\documentclass[journal,twoside]{IEEEtran}
\usepackage{graphicx,cite,amssymb,amsmath,psfrag,subfigure}
\usepackage{graphicx}
\usepackage{amssymb}
\usepackage{amsmath}
\usepackage{cite}
\usepackage{subfigure}
\usepackage{mathrsfs}
\usepackage[displaymath,mathlines]{lineno}
\usepackage{color}
\usepackage{tabulary}
\usepackage{multirow}

\newtheorem{theorem}{Theorem}

\newtheorem{proposition}{Proposition}

\newtheorem{remark}{Remark}
\newtheorem{algorithm}{Algorithm}

\DeclareMathOperator{\E}{\mathtt{E}}

\DeclareMathOperator{\tr}{\mathrm{tr}}
\DeclareMathOperator{\var}{\mathbb{V}\mathrm{ar}}

\newcommand{\Nt}{N_{\mathrm{tx}}}
\newcommand{\Nr}{N_{\mathrm{rx}}}

\newcommand{\CG}[2]{\mathcal{CN}\left({#1},{#2}\right)}

\newcommand{\B}[1]{{\mathbf{#1}}}

\newcommand{\Pp}{p_{\mathrm{p}}}

\newcommand{\ER}{E_{\mathsf{R}}}
\newcommand{\ES}{E_{\mathsf{S}}}

\newcommand{\Ps}{p_{\tt{S}}}
\newcommand{\Psk}{p_{{\tt S},k}}
\newcommand{\PR}{p_{\tt{R}}}

\makeatletter
\def\@setsize#1#2#3#4{
    \@nomath#1
    \let\@currsize#1
    \baselineskip #2
    \baselineskip \baselinestretch\baselineskip
    \parskip \baselinestretch\parskip
    \setbox\strutbox \hbox{
        \vrule height.7\baselineskip
            depth.3\baselineskip
            width\z@}
    \skip\footins \baselinestretch\skip\footins
    \normalbaselineskip\baselineskip#3#4}
\makeatother

\makeatletter
\newcommand{\setstretch}[1]{
    \def\baselinestretch{#1}%
    \@currsize
    }
\makeatother



\def\BibTeX{{\rm B\kern-.05em{\sc i\kern-.025em b}\kern-.08em
    T\kern-.1667em\lower.7ex\hbox{E}\kern-.125emX}}

\newcounter{eqncnt}

\newcounter{eqnback}

\begin{document}
\vspace{0 cm}
\title{
    Multipair Full-Duplex Relaying with Massive Arrays and Linear Processing}
\author{
        Hien Quoc Ngo, Himal A. Suraweera, Michail Matthaiou, and
        Erik G. Larsson\vspace{-0.0cm}
\thanks{
        H.~Q.\ Ngo and E.~G.\ Larsson are with the Department of Electrical
    Engineering (ISY), Link\"{o}ping University, 581 83 Link\"{o}ping,
    Sweden
        (email: nqhien@isy.liu.se; egl@isy.liu.se).
}
\thanks{
        H. A. Suraweera is with the Department of Electrical \& Electronic Engineering, University
of Peradeniya, Peradeniya 20400, Sri Lanka (email:
himal@ee.pdn.ac.lk). }
\thanks{
M. Matthaiou is with the School of Electronics, Electrical
Engineering and Computer Science, Queen's University Belfast,
Belfast, BT3 9DT, U.K., and with the Department of Signals and
Systems, Chalmers University of Technology, 412 96 Gothenburg,
Sweden (email: m.matthaiou@qub.ac.uk).}
\thanks{The work of H.~Q.\ Ngo and E.~G.\ Larsson was supported  in part by the Swedish Research Council
(VR), the Swedish
    Foundation for Strategic Research (SSF), and ELLIIT.}
\thanks{Part of this work will be presented at the 2014 IEEE International Conference on Communications (ICC) \cite{NSML:14:ICC}.}
}

\markboth{}
        {}

\maketitle

\vspace{-1.2cm}
\begin{abstract}
We consider a multipair decode-and-forward relay channel, where
multiple sources transmit simultaneously their signals to multiple
destinations with the help of a full-duplex relay station. We
assume that the relay station is equipped with massive arrays,
while all sources and destinations have a single antenna. The
relay station uses channel estimates obtained from received pilots
and zero-forcing (ZF) or maximum-ratio combining/maximum-ratio
transmission (MRC/MRT) to process the signals. To reduce
significantly the loop interference effect, we propose two
techniques: i) using a massive receive antenna array; or ii) using
a massive transmit antenna array together with very low transmit
power at the relay station. We derive an exact achievable rate in
closed-form for MRC/MRT processing and an analytical approximation
of the achievable rate for ZF processing. This approximation is
very tight, especially for large number of relay station antennas.
These closed-form expressions enable us to determine the regions
where the full-duplex mode outperforms the half-duplex mode, as
well as, to design an optimal power allocation scheme. This
optimal power allocation scheme aims to maximize the energy
efficiency for a given sum spectral efficiency and under peak
power constraints at the relay station and sources. Numerical
results verify the effectiveness of the optimal power allocation
scheme. Furthermore, we show that, by doubling the number of
transmit/receive antennas at the relay station, the transmit power
of each source and of the relay station can be reduced by $1.5$dB
if the pilot power is equal to the signal power, and by $3$dB if
the pilot power is kept fixed, while maintaining a given
quality-of-service.
\end{abstract}


\section{Introduction} \label{Sec:Introduction}
Multiple-input multiple-output (MIMO) systems that use antenna
arrays with a few hundred antennas for multiuser operation
(popularly called ``Massive MIMO'') is an emerging technology that
can deliver all the attractive benefits of traditional MIMO, but
at a much larger scale
\cite{Mar:10:WCOM,RPLLMET:11:SPM,Eri:13:MCOM}. Such systems can
reduce substantially the effects of noise, fast fading and
interference and provide increased throughput. Importantly, these
attractive features of massive MIMO can be reaped using simple
signal processing techniques and at a reduction of the total
transmit power. As a result, not surprisingly, massive MIMO
combined with cooperative relaying is a strong candidate for the
development of future energy-efficient cellular networks
\cite{Eri:13:MCOM,SNDYL:13:ICC}.

On a parallel avenue, full-duplex  relaying has received a lot of
research interest, for its ability to recover the bandwidth loss
induced by conventional half-duplex relaying. With full-duplex
relaying, the relay node receives and transmits simultaneously on
the same channel \cite{BLISS1,BLISS2}. As such, full-duplex
utilizes the spectrum resources more efficiently. Over the recent
years, rapid progress has been made on both theory and
experimental hardware platforms to make full-duplex wireless
communication an efficient practical solution
\cite{RWW:11:SP,RWW:11:WCOM,RWW:11:ACSSC,ZHENG1,AKSRC:12:MobiCom,DUARTE1}.
The benefit of improved spectral efficiency in the full-duplex
mode comes at the price of loop interference  due to signal
leakage from the relay's output to the input
\cite{RWW:11:WCOM,RWW:11:ACSSC}. A large amplitude difference
between the loop interference and the received signal coming from
the source can exceed the dynamic range of the analog-to-digital
converter at the receiver side, and, thus, its mitigation is
crucial for full-duplex operation \cite{DUARTE1,DUARTE2}. Note
that how to overcome the detrimental effects of loop interference
is a highly active area in full-duplex research.

Traditionally, loop interference suppression is performed in the
antenna domain using a variety of passive techniques that
electromagnetically shield the transmit antenna from the receive
antenna. As an example, directional antennas can be used to place
a null at the receive antenna. Since the distance between the
transmit and receive arrays is short, such techniques require
significant levels of loop interference mitigation and, hence, are
hard to realize. On the other hand, active time domain loop
interference cancellation techniques use the knowledge of the
interfering signal to pre-cancel the loop interference in the
radio frequency signal and achieve higher levels of loop
interference suppression. However, they demand advanced noise
cancellation methods and sophisticated electronic implementation
\cite{RWW:11:SP}. Yet, MIMO processing provides an effective means
of suppressing the loop interference in the spatial domain. With
multiple transmit or receive antennas at the full-duplex relay,
precoding solutions, such as zero-forcing (ZF), can be deployed to
mitigate the loop interference effects. Although sub-optimal in
general, simple ZF-based precoder can completely cancel the loop
interference and remove the closed-loop between the relay's input
and output. Several papers have considered spatial loop
interference suppression; for example, \cite{RWW:11:ACSSC}
proposes to direct the loop interference of a full-duplex
decode-and-forward (DF) relay to the least harmful spatial
dimensions. In \cite{RWW:11:SP}, assuming a multiple antenna
relay, a range of spatial suppression techniques including
precoding and antenna selection is analyzed. 
In \cite{YSUNG}, several antenna sub-set selection schemes are
proposed aiming to suppress loop interference at the relay's
transmit side. More recently, \cite{HIMAL} analyzed several
antenna selection schemes for spatial loop interference
suppression in a MIMO relay channel.

Different from the majority of existing works in the literature,
which consider systems that deploy only few antennas, in this
paper we consider a massive MIMO full-duplex relay architecture.
The large number of spatial dimensions available in a massive MIMO
system can be effectively used to suppress the loop interference
in the spatial domain. We assume that a group of $K$ sources
communicate with a group of $K$ destinations using a massive MIMO
full-duplex relay station. Specifically, in this multipair massive
MIMO relay system, we deploy two processing schemes, namely, ZF
and maximum ratio combining (MRC)/maximal ratio transmission (MRT)
with full-duplex relay operation. Recall that linear processing
techniques, such as ZF or MRC/MRT processing, are low-complexity
solutions that are anticipated to be utilized in massive MIMO
topologies. Their main advantage is that in the large-antenna
limit, they can perform as well as non-linear schemes (e.g.,
maximum-likelihood) \cite{Mar:10:WCOM,SNDYL:13:ICC,ZHANG-COM-MAG}.
Our system setup could be applied in cellular networks, where
several users transmit simultaneously signals to several other
users with the help of a relay station (infrastructure-based
relaying). Note that, newly evolving wireless standards, such as
LTE-Advanced, promote the use of relays (with unique cell ID and
right for radio resource management) to serve as low power base
stations \cite{ZDPWL:11:WCOM,YHXM:09:CMAG}.

We investigate the achievable rate and power efficiency of the
aforementioned full-duplex system setup. Moreover, we compare
full-duplex and half-duplex modes and show the benefit of choosing
one over the other (depending on the loop interference level of
the full-duplex mode). Although the current work uses techniques
related to those in Massive MIMO, we investigate a substantially
different setup. Specifically, previous works related to Massive
MIMO systems
\cite{Mar:10:WCOM,RPLLMET:11:SPM,Eri:13:MCOM,NLM:TCOM:13}
considered the uplink or the downlink of multiuser MIMO channels.
In contrast, we consider multipair full-duplex relaying channels
with massive arrays at the relay station. As a result, our new
contributions are very different from the existing works on
Massive MIMO. The main contributions of this paper are summarized
as follows:

\begin{enumerate}
    \item We show that the loop interference can be significantly reduced,  if the relay station is
equipped with a large receive antenna array or/and is equipped
with a large transmit antenna array. At the same time, the
inter-pair interference and noise effects disappear. Furthermore,
when the number of relay station transmit antennas, $\Nt$, and the
number of relay station receive antennas, $\Nr$, are large, we can
scale down the transmit powers of each source and of the relay
proportionally to $1/\Nr$ and $1/\Nt$, respectively, if the pilot
power is kept fixed, and proportionally to $1/\sqrt{\Nr}$ and
$1/\sqrt{\Nt}$, respectively, if the pilot power and the data
power are the same.

    \item We derive exact and approximate closed-form expressions for the end-to-end (e2e) achievable rates of MRC/MRT and  ZF processing, respectively.
    These simple closed-form expressions enable us to  obtain
    important insights as well as to compare full-duplex and half-duplex operation and demonstrate which mode
yields better performance. As a general remark, the full-duplex
mode  improves significantly the overall system performance when
the loop interference level is low. In addition, we propose the
use of a hybrid mode for each large-scale fading realization,
which switches between the full-duplex and half-duplex modes,  to
maximize the sum spectral efficiency.

    \item We design an optimal power allocation  algorithm for the data transmission phase, which maximizes the energy efficiency for a desired sum spectral
    efficiency and under peak power constraints at the relay station and sources. This optimization problem can be approximately solved  via a sequence of geometric programs
    (GPs). Our numerical results indicate that the proposed power allocation improves notably the performance compared
    to uniform power allocation.

    \end{enumerate}

\textit{Notation:} We use boldface upper- and lower-case letters
to denote matrices and column vectors, respectively. The
superscripts $()^\ast$, $()^T$, and $()^H$ stand for the
conjugate, transpose, and conjugate-transpose, respectively. The
Euclidean norm, the trace, the expectation, and the variance
operators are denoted by $\|\cdot\|$, $\tr\left(\cdot\right)$,
$\mathbb{E}\left\{\cdot\right\}$, and $\var\left(\cdot \right)$,
respectively. The notation $\mathop \to \limits^{a.s.}$ means
almost sure convergence, while  $\mathop \to \limits^{d}$ means
convergence in distribution. Finally, we use $\B{z} \sim
\CG{\mathbf{0}}{\B{\Sigma}}$ to denote a circularly symmetric
complex Gaussian vector $\B{z}$ with zero mean and covariance
matrix $\B{\Sigma}$.

\begin{figure}[t]
    \centerline{\includegraphics[width=0.48\textwidth]{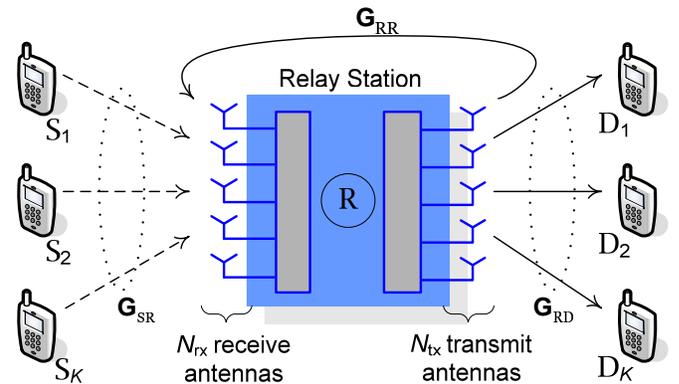}}
    \caption{Multipair full-duplex relaying system.}
    \label{fig:1}
\end{figure}

\section{System Model} \label{Sec:SysModel}

Figure~\ref{fig:1} shows the considered multipair DF relaying
system where $K$ communication pairs $({\tt S}_k, {\tt D_k})$,
$k=1, \ldots, K$, share the same time-frequency resource and a
common relay station, ${\tt R}$. The $k$th source, ${\tt S}_k$,
communicates with the $k$th destination, ${\tt D}_k$, via the
relay station, which operates in a full-duplex mode. All source
and destination nodes are equipped with a single antenna, while
the relay station is equipped with $\Nr$ receive antennas and
$\Nt$ transmit antennas. The total number of antennas at the relay
station is $N=\Nr+\Nt$. We assume that the hardware chain
calibration is perfect so that the channel from the relay station
to the destination is reciprocal \cite{Eri:13:MCOM}. Further, the
direct links among ${\tt S}_k$ and ${\tt D_k}$ do not exist due to
large path loss and heavy shadowing. Our network configuration is
of practical interest, for example, in a cellular setup, where
inter-user communication is realized with the help of a base
station equipped with massive arrays.

At time instant $i$, all $K$ sources ${\tt S}_k$, $k=1, ..., K$,
 transmit simultaneously their signals, $\sqrt{\Ps}x_k\left[
i\right]$, to the relay station, while the relay station
broadcasts $\sqrt{\PR}\B{s}\left[i\right] \in
\mathbb{C}^{\Nt\times 1}$ to all $K$ destinations. Here, we assume
that $\E\left\{\left|x_k\left[ i\right]\right|^2\right\}=1$ and
$\E\left\{\left\|\B{s}\left[i\right]\right\|^2\right\}=1$ so that
$\Ps$ and $\PR$ are the average transmit powers of each source and
of the relay station. Since the relay station receives and
transmits at the same frequency, the received signal at the relay
station is interfered by its own transmitted signal,
$\B{s}\left[i\right]$. This is called \emph{loop interference}.
Denote by $\B{x}\left[i\right]\triangleq \left[x_1\left[ i\right]
~ x_2\left[ i\right] ~ ... ~ x_K\left[ i\right] \right]^T$. The
received signals at the relay station and the $K$ destinations are
given by \cite{RWW:11:SP}
\begin{align} \label{eq Sys 1}
    \B{y}_{\tt R}\left[i \right]
    &=
    \sqrt{\Ps}
    \B{G}_{\tt SR}
    \B{x}\left[i\right]
    +
    \sqrt{\PR}
    \B{G}_{\tt RR}
    \B{s}\left[i\right]
    +
    \B{n}_{\tt R}\left[i\right],
    \\
    \B{y}_{\tt{D}}\left[i \right]
    &=
    \sqrt{\PR}
    \B{G}_{\tt RD}^T
    \B{s}\left[i\right]
    +
    \B{n}_{\tt D}\left[i\right], \label{eq Sys 1b}
\end{align}
respectively, where $\B{G}_{\tt SR} \in \mathbb{C}^{\Nr\times K}$
and $\B{G}_{\tt RD}^T \in \mathbb{C}^{K \times \Nt}$ are the
channel matrices from the $K$ sources to the relay station's
receive antenna array and from the relay station's transmit
antenna array to the $K$ destinations, respectively. The channel
matrices account for both small-scale fading and large-scale
fading. More precisely, $\B{G}_{\tt SR}$ and $\B{G}_{\tt RD}$ can
be expressed as $\B{G}_{\tt SR}=\B{H}_{\tt SR} \B{D}_{\tt
SR}^{1/2}$ and $\B{G}_{\tt RD}=\B{H}_{\tt RD} \B{D}_{\tt
RD}^{1/2}$, where the small-scale fading matrices $\B{H}_{\tt SR}$
and $\B{H}_{\tt RD}$ have independent and identically distributed
(i.i.d.) $\CG{0}{1}$ elements, while $\B{D}_{\tt SR}$ and
$\B{D}_{\tt RD}$ are the large-scale fading diagonal matrices
whose $k$th diagonal elements are denoted by $\beta_{{\tt SR},k}$
and $\beta_{{\tt RD},k}$, respectively. The above channel models
rely on the favorable propagation assumption, which assumes that
the channels from the relay station to different sources and
destinations are independent \cite{Eri:13:MCOM}. The validity of
this assumption was demonstrated in practice, even for massive
arrays \cite{GERT:11:VTC}. Also in \eqref{eq Sys 1}, $\B{G}_{\tt
RR} \in \mathbb{C}^{\Nr\times \Nt}$ is the channel matrix between
the transmit and receive arrays which represents the loop
interference. We model the loop interference channel via the
Rayleigh fading distribution, under the assumptions that any
line-of-sight component is efficiently reduced by antenna
isolation and the major effect comes from scattering. Note that if
hardware loop interference cancellation is applied, $\B{G}_{\tt
RR}$ represents the residual interference due to imperfect loop
interference cancellation. The residual interfering link is also
modeled as a Rayleigh fading channel, which is a common assumption
made in the existing literature \cite{RWW:11:SP}. Therefore, the
elements of $\B{G}_{\tt RR}$ can be modeled as i.i.d.
$\CG{0}{\sigma_{\tt LI}^2}$ random variables, where $\sigma_{\tt
LI}^2$ can be understood as the level of loop interference, which
depends on the distance between the transmit and receive antenna
arrays or/and the capability of the hardware loop interference
cancellation technique \cite{RWW:11:WCOM}. Here, we assume that
the distance between the transmit array and the receive array is
much larger than the inter-element distance, such that the
channels between the transmit and receive antennas are
i.i.d.;\footnote{For example, consider two transmit and receive
arrays which are located on the two sides of a building with a
distance of $3$m. Assume that the system is operating at 2.6GHz.
Then, to guarantee uncorrelation between the antennas, the
distance between adjacent antennas is about $6$cm, which is half a
wavelength. Clearly, $3$m $\gg$ $6$cm. In addition, if each array
is a cylindrical array with 128 antennas, the physical size of
each array is about $28$cm $\times 29$cm \cite{GERT:11:VTC} which
is still relatively small compared to the distance between the two
arrays.} also, $\B{n}_{\tt R}\left[i\right]$ and $\B{n}_{\tt
D}\left[i\right]$ are additive white Gaussian noise (AWGN) vectors
at the relay station and the $K$ destinations, respectively. The
elements of $\B{n}_{\tt R}\left[i\right]$  and $\B{n}_{\tt
D}\left[i\right]$ are assumed to be i.i.d.\ $\CG{0}{1}$.

\subsection{Channel Estimation}

In practice, the channels ${\B{G}}_{\tt SR}$ and ${\B{G}}_{\tt
RD}$ have to be estimated at the relay station. 
The standard way of doing this is to utilize pilots
\cite{Mar:10:WCOM}. To this end, a part of the coherence interval
is used for channel estimation. All sources and destinations
transmit simultaneously their pilot sequences of $\tau$ symbols to
the relay station. The received pilot matrices at the relay
receive and transmit antenna arrays are given by
\begin{align} \label{eq CE 1}
    \B{Y}_{\mathrm{rp}}
    &=
    \sqrt{\tau \Pp}
    \B{G}_{\tt SR}
    \B{\Phi}_{\tt S}
    +
    \sqrt{\tau \Pp}
    \bar{\B{G}}_{\tt RD}
    \B{\Phi}_{\tt D}
    +
    \B{N}_{\mathrm{rp}},
    \\\label{eq CE 1b}
    \B{Y}_{\mathrm{tp}}
    &=
    \sqrt{\tau \Pp}
    \bar{\B{G}}_{\tt SR}
    \B{\Phi}_{\tt S}
    +
    \sqrt{\tau \Pp}
    \B{G}_{\tt RD}
    \B{\Phi}_{\tt D}
    +
    \B{N}_{\mathrm{tp}},
\end{align}
respectively, where $\bar{\B{G}}_{\tt SR} \in
\mathbb{C}^{\Nt\times K}$ and $\bar{\B{G}}_{\tt RD} \in
\mathbb{C}^{\Nr \times K}$ are the channel matrices from the $K$
sources to the relay station's transmit antenna array and from the
$K$ destinations to the relay station's receive antenna array,
respectively; $\Pp$ is the transmit power of each pilot symbol,
$\B{N}_{\mathrm{rp}}$ and $\B{N}_{\mathrm{tp}}$ are  AWGN matrices
which include i.i.d. $\CG{0}{1}$ elements, while the $k$th rows of
$\B{\Phi}_{\tt S} \in \mathbb{C}^{K\times\tau}$ and $\B{\Phi}_{\tt
D}\in \mathbb{C}^{K\times\tau}$ are the pilot sequences
transmitted from ${\tt S}_k$ and ${\tt D}_k$, respectively. All
pilot sequences are assumed to be pairwisely orthogonal, i.e.,
$\B{\Phi}_{\tt S} \B{\Phi}_{\tt S}^H = \B{I}_K$, $\B{\Phi}_{\tt D}
\B{\Phi}_{\tt D}^H = \B{I}_K$, and $\B{\Phi}_{\tt S} \B{\Phi}_{\tt
D}^H = \B{0}_K$. This requires that $\tau \geq 2K$.

We assume that the relay station uses minimum mean-square-error
(MMSE) estimation to estimate $\B{G}_{\tt SR}$ and $\B{G}_{\tt
RD}$. The MMSE channel estimates of $\B{G}_{\tt SR}$ and
$\B{G}_{\tt RD}$ are given by \cite{KAY:93:Book}
\begin{align} \label{eq MMSECE 1}
    \hat{\B{G}}_{\tt SR}
    =
    \frac{1}{\sqrt{\tau \Pp} }
    \B{Y}_{\mathrm{rp}} \B{\Phi}_{\tt S}^H \tilde{\B{D}}_{\tt SR}
    =
    \B{G}_{\tt SR} \tilde{\B{D}}_{\tt SR}
    +
    \frac{1}{\sqrt{\tau \Pp} }
    \B{N}_{\tt S}\tilde{\B{D}}_{\tt SR},
    \\
    \hat{\B{G}}_{\tt RD}
    =
    \frac{1}{\sqrt{\tau \Pp} }
    \B{Y}_{\mathrm{tp}} \B{\Phi}_{\tt D}^H \tilde{\B{D}}_{\tt RD}
    =
    \B{G}_{\tt RD} \tilde{\B{D}}_{\tt RD}
    +
    \frac{1}{\sqrt{\tau \Pp} }
    \B{N}_{\tt D}\tilde{\B{D}}_{\tt RD},
\end{align}
respectively, where $\tilde{\B{D}}_{\tt SR} \triangleq
\left(\frac{{\B{D}}_{\tt SR}^{-1}}{\tau \Pp} + \B{I}_K
\right)^{-1}$, $\tilde{\B{D}}_{\tt RD} \triangleq
\left(\frac{{\B{D}}_{\tt RD}^{-1}}{\tau \Pp} + \B{I}_K
\right)^{-1}$, $\B{N}_{\tt S} \triangleq \B{N}_{\mathrm{rp}}
\B{\Phi}_{\tt S}^H$ and $\B{N}_{\tt D} \triangleq
\B{N}_{\mathrm{tp}} \B{\Phi}_{\tt D}^H$. Since the rows of
$\B{\Phi}_{\tt S}$ and $\B{\Phi}_{\tt D}$ are pairwisely
orthogonal, the elements of $\B{N}_{\tt S}$ and $\B{N}_{\tt D}$
are i.i.d. $\CG{0}{1}$ random variables. Let
$\pmb{\mathcal{E}}_{\tt SR}$ and $\pmb{\mathcal{E}}_{\tt RD}$ be
the estimation error matrices of $\B{G}_{\tt SR}$ and $\B{G}_{\tt
RD}$, respectively. Then,
\begin{align} \label{eq MMSECE 2}
    \B{G}_{\tt SR}
    &=
    \hat{\B{G}}_{\tt SR}
    +
    \pmb{\mathcal{E}}_{\tt SR},
    \\
    \B{G}_{\tt RD}
    &=
    \hat{\B{G}}_{\tt RD}
    +
    \pmb{\mathcal{E}}_{\tt RD}. \label{eq MMSECE 2b}
\end{align}
From the property of MMSE channel estimation, $\hat{\B{G}}_{\tt
SR}$, $\pmb{\mathcal{E}}_{\tt SR}$, $\hat{\B{G}}_{\tt RD}$, and
$\pmb{\mathcal{E}}_{\tt RD}$ are independent \cite{KAY:93:Book}.
Furthermore, we have that the rows of $\hat{\B{G}}_{\tt SR}$,
$\pmb{\mathcal{E}}_{\tt SR}$, $\hat{\B{G}}_{\tt RD}$, and
$\pmb{\mathcal{E}}_{\tt RD}$ are mutually independent and
distributed as $\CG{\B{0}}{\hat{\B{D}}_{{\tt SR}}}$,
$\CG{\B{0}}{{\B{D}}_{{\tt SR}}-\hat{\B{D}}_{{\tt SR}}}$,
$\CG{\B{0}}{\hat{\B{D}}_{{\tt RD}}}$, and
$\CG{\B{0}}{{\B{D}}_{{\tt RD}}-\hat{\B{D}}_{{\tt RD}}}$,
respectively, where $\hat{\B{D}}_{{\tt SR}}$ and
$\hat{\B{D}}_{{\tt RD}}$ are diagonal matrices whose $k$th
diagonal elements are $\sigma_{{\tt SR},k}^2\triangleq \frac{\tau
\Pp \beta_{{\tt SR},k}^2}{\tau \Pp \beta_{{\tt SR},k} +1}$ and
 $\sigma_{{\tt RD},k}^2\triangleq \frac{\tau \Pp \beta_{{\tt
RD},k}^2}{\tau \Pp \beta_{{\tt RD},k} +1}$, respectively.

\subsection{Data Transmission}
The relay station considers the channel estimates as the true
channels and employs linear processing. More precisely, the relay
station uses a linear receiver to decode the signals transmitted
from the $K$ sources. Simultaneously, it uses a linear precoding
scheme to forward the signals to the $K$ destinations.

\subsubsection{Linear Receiver}
With the linear receiver, the received signal $\B{y}_{\tt
R}\left[i \right]$ is separated into $K$ streams by multiplying it
with a linear receiver matrix $\B{W}^T$ (which is a function of
the channel estimates) as follows:
\begin{align} \label{eq ZFR 1}
    \B{r}\left[i \right]
    &=
    \B{W}^T \B{y}_{\tt R}\left[i \right]
    =
    \sqrt{\Ps}
    \B{W}^T
    \B{G}_{\tt SR}
    \B{x}\left[i\right]
    +
    \sqrt{\PR}
    \B{W}^T
    \B{G}_{\tt RR}
    \B{s}\left[i\right]
    +
    \B{W}^T
    \B{n}_{\tt R}\left[i\right].
\end{align}
Then, the $k$th stream ($k$th element of $\B{r}\left[i \right]$)
is used to decode the signal transmitted from ${\tt S}_k$. The
$k$th element of $\B{r}\left[i \right]$ can be expressed as
\begin{align} \label{eq ZFR 1b}
    {r}_k\left[i \right]
    &=
    \underbrace
    {\sqrt{\Ps}
    \B{w}_k^T
    \B{g}_{{\tt SR},k}
    {x}_k\left[i\right]}_{\text{desired signal}}
    +
    \underbrace
    {\sqrt{\Ps}
    \sum_{j\neq k}^K
    \B{w}_k^T
    \B{g}_{{\tt SR},j}
    {x}_j\left[i\right]}_{\text{interpair
    interference}}\nonumber\\
    &+
    \underbrace
    {\sqrt{\PR}
    \B{w}_k^T
    \B{G}_{\tt RR}
    \B{s}\left[i\right]}_{\text{loop interference}}
    +
    \underbrace
    {\B{w}_k^T
    \B{n}_{\tt R}\left[i\right]}_{\text{noise}},
\end{align}
where $\B{g}_{{\tt SR},k}$, $\B{w}_k$ are the $k$th columns of
$\B{G}_{\tt SR}$, $\B{W}$, respectively, and ${x}_k\left[i\right]$
is the $k$th element of $\B{x}\left[i\right]$.

\subsubsection{Linear Precoding}
After detecting the signals transmitted from the $K$ sources, the
relay station uses  linear precoding to process these signals
before broadcasting them to all $K$ destinations. Owing to the
processing delay \cite{RWW:11:SP}, the transmit vector
$\B{s}\left[i\right]$ is a precoded version of $\B{x}\left[i -
d\right]$, where $d$ is the processing delay. More precisely,
\begin{align} \label{eq ZFP 1}
    \B{s}\left[i \right]
    &=
    \B{A}\B{x}\left[i - d\right],
\end{align}
where $\B{A} \in \mathbb{C}^{\Nt\times K}$ is a linear precoding
matrix which is a function of the channel estimates. We assume
that the processing delay $d\geq 1$ which guarantees that the
receive and transmit signals at the relay station, for a given
time instant, are uncorrelated. This is a common assumption for
full-duplex systems in the existing literature
\cite{RWW:11:WCOM,ZHENG1}.

From \eqref{eq Sys 1b} and \eqref{eq ZFP 1}, the received signal
at ${\tt D}_k$ can be expressed as
\begin{align} \label{eq ZFR 3}
    {y}_{{\tt D},k}\left[i \right]
    &=
    \sqrt{\PR}
    \B{g}_{{\tt RD},k}^T
    \B{a}_k {x}_k\left[i-d\right]\nonumber\\
    &+
    \sqrt{\PR}
    \sum_{j\neq k}^K
    \B{g}_{{\tt RD},k}^T
    \B{a}_j {x}_j\left[i-d\right]
    +
    {n}_{{\tt D},k}\left[i\right],
\end{align}
where $\B{g}_{{\tt RD},k}$, $\B{a}_k$ are the $k$th columns of
$\B{G}_{\tt RD}$, $\B{A}$, respectively, and ${n}_{{\tt
D},k}\left[i\right]$ is the $k$th element of $\B{n}_{{\tt
D}}\left[i\right]$.

\subsection{ZF and MRC/MRT Processing}
In this work, we consider two common linear processing techniques:
ZF and MRC/MRT processing.

\subsubsection{ZF Processing}

In this case, the relay station uses the ZF receiver and ZF
precoding to process the signals. Due to the fact that all
communication pairs share the same time-frequency resource, the
transmission of a given pair will be impaired by the transmissions
of other pairs. This effect is called ``interpair interference''.
More explicitly, for the transmission from ${\tt S}_k$ to the
relay station, the interpair interference is represented by the
term $\sqrt{\Ps}
    \sum_{j\neq k}^K
    \B{w}_k^T
    \B{g}_{{\tt SR},j}
    {x}_j\left[i\right]$, while for the transmission from the relay station to ${\tt D}_k$, the interpair interference is $    \sqrt{\PR}
    \sum_{j\neq k}^K
    \B{g}_{{\tt RD},k}^T
    \B{a}_j {x}_j\left[i-d\right]$. With ZF processing, interpair
interference is nulled out by projecting each stream onto the
orthogonal complement of the interpair interference. This can be
done if the relay station has perfect channel state information
(CSI). However, in practice, the relay station knows only the
estimates of CSI. Therefore, interpair interference and loop
interference still exist. We assume that $\Nr, \Nt > K$.

The ZF receiver and ZF precoding matrices are respectively given
by \cite{NLM:TCOM:13,YM:13:JSAC}
\begin{align}\label{eq ZFReceiver 1}
    \B{W}^T
    &=
    \B{W}_{\tt ZF}^T
        \triangleq
        \left(\hat{\B{G}}_{\tt SR}^H \hat{\B{G}}_{\tt SR}\right)^{-1}\hat{\B{G}}_{\tt
        SR}^H,
    \\
\label{eq ZFPrecoder 1}
    \B{A}
    &=
    \B{A}_{\tt ZF}
    \triangleq
    \alpha_{\tt ZF} \hat{\B{G}}_{\tt RD}^{\ast}
    \left(
        \hat{\B{G}}_{\tt RD}^T \hat{\B{G}}_{\tt RD}^{\ast}
    \right)^{-1}\!\!\!,
\end{align}
where $\alpha_{\tt ZF}$ is a normalization constant, chosen to
satisfy a long-term total transmit power constraint at the relay,
i.e., $\E\left\{ \left\| \B{s}\left[ i\right] \right\|^2 \right\}
= 1$. Therefore, we have \cite{YM:13:JSAC}
\begin{align} \label{eq ZFR 2}
    \alpha_{\tt ZF}
    &=
    \sqrt{
        \frac{
            \Nt-K
            }{
            \sum_{k=1}^K \sigma_{{\tt RD},k}^{-2}
            }
    }.
\end{align}
%
\subsubsection{MRC/MRT Processing}

The ZF processing neglects the effect of noise and, hence, it
works poorly when the signal-to-noise ratio (SNR) is low. By
contrast, the MRC/MRT processing aims to maximize the received
SNR, by neglecting the interpair interference effect. Thus,
MRC/MRT processing works well at low SNRs, and works poorly at
high SNRs. With MRC/MRT processing, the relay station uses MRC to
detect the signals transmitted from the $K$ sources. Then, it uses
the MRT technique to transmit signals towards the $K$
destinations. The MRC receiver and MRT precoding matrices are
respectively given by \cite{NLM:TCOM:13,YM:13:JSAC}
\begin{align}\label{eq MRCReceiver 1}
    \B{W}^T
    &=\B{W}_{\tt MRC}^T
        \triangleq
        \hat{\B{G}}_{\tt SR}^H,
    \\
\label{eq MRTPrecoder 1}
    \B{A}
    &=
    \B{A}_{\tt MRT}
    \triangleq
    \alpha_{\tt MRT} \hat{\B{G}}_{\tt RD}^{\ast},
\end{align}
where the normalization constant  $\alpha_{\tt MRT}$ is  chosen to
satisfy a long-term total transmit power constraint at the relay,
i.e., $\E\left\{ \left\| \B{s}\left[ i\right] \right\|^2 \right\}
= 1$, and we have \cite{YM:13:JSAC}
\begin{align} \label{eq MRT 2}
    \alpha_{\tt MRT}
    &=
    \sqrt{
        \frac{
            1
            }{
            \Nt \sum_{k=1}^K\sigma_{{\tt RD},k}^2
            }
    }.
\end{align}

\section{Loop Interference Cancellation with Large Antenna Arrays}

In this section, we consider the potential of using massive MIMO
technology to cancel the loop interference due to the full-duplex
operation at the relay station. Some interesting insights are also
presented.\vspace{-0.5cm}

\subsection{Using a Large Receive Antenna Array ($\Nr\to\infty$)}\label{subsec large1}
The loop interference can be canceled out by projecting it onto
its orthogonal complement. However, this orthogonal projection may
harm the desired signal. Yet, when $\Nr$ is large, the subspace
spanned by the loop interference is nearly orthogonal to the
desired signal's subspace and, hence, the orthogonal projection
scheme will perform very well. The next question is how to project
the loop interference component? It is interesting to observe
that, when $\Nr$ grows large, the channel vectors of the desired
signal and the loop interference become nearly orthogonal.
Therefore, the ZF or the MRC receiver can act as an orthogonal
projection of the loop interference. As a result, the loop
interferenceI can be reduced significantly by using large $\Nr$
together with the ZF or MRC receiver. This observation is
summarized in the following proposition.

\begin{proposition}\label{PropLarge1}
Assume that the number of source-destination pairs, $K$, is fixed.
For any finite $\Nt$ or for any $\Nt$, such that $\Nr/\Nt$ is
fixed, as $\Nr \to \infty$, the received signal at the relay
station for decoding the signal transmitted from ${\tt S}_k$  is
given by
\begin{align} \label{eq PropLarge 1}
    {r}_k\left[i \right]
    &\mathop  \to \limits^{a.s.}
    \sqrt{\Ps}
    {x}_k\left[i\right], ~ \text{for ZF},
    \\
    \frac{{r}_k\left[i \right]}{\Nr\sigma_{{\tt SR}, k}^2}
    &\mathop  \to \limits^{a.s.}
    \sqrt{\Ps}
    {x}_k\left[i\right], ~ \text{for MRC/MRT}. \label{eq PropLarge 1b}
\end{align}
\begin{proof}
See Appendix~\ref{Appendix Large1}.
\end{proof}
\end{proposition}
The aforementioned results imply that, when $\Nr$ grows to
infinity, the loop interference can be canceled out. Furthermore,
the interpair interference and noise effects also disappear. The
received signal at the relay station after using ZF or MRC
receivers includes only the desired signal and, hence, the
capacity of the communication link ${\tt S}_k \to {\tt R}$ grows
without bound. As a result, the system performance is limited only
by the performance of the communication link ${\tt R} \to {\tt
D}_k$ which does not depend on the loop interference.

\subsection{Using a Large Transmit Antenna Array and Low Transmit Power ($\PR=\ER/\Nt$, where $\ER$ is Fixed, and  $\Nt \to
\infty$)}\label{subsec large2}

The loop interference depends strongly on the transmit power at
the relay station, $\PR$ and, hence, another way to reduce it is
to use low transmit power $\PR$. Unfortunately, this will also
reduce the quality of the transmission link ${\tt R} \to {\tt
D}_k$ and, hence, the e2e system performance will be degraded.
However, with a large relay station transmit antenna array, we can
reduce the relay transmit power while maintaining a desired
quality-of-service (QoS) of the transmission link ${\tt R} \to
{\tt D}_k$. This is due to the fact that, when the number of
transmit antennas, $\Nt$, is large, the relay station can focus
its emitted energy into the physical directions wherein the
destinations are located. At the same time, the relay station can
purposely avoid transmitting into physical directions where the
receive antennas are located and, hence, the loop interference can
be significantly reduced. Therefore, we propose to use a very
large $\Nt$ together with low transmit power at the relay station.
With this method, the loop interference in the transmission link
${\tt S}_k \to {\tt R}$ becomes negligible, while the quality of
the transmission link ${\tt R} \to {\tt D}_k$ is still fairly
good. As a result, we can obtain a good e2e performance.

\begin{proposition}\label{PropLarge2}
Assume that $K$ is fixed and the transmit power at the relay
station is $\PR=\ER/\Nt$, where $\ER$ is fixed regardless of
$\Nt$. For any finite $\Nr$, as $\Nt \to \infty$, the received
signals at the relay station and ${\tt D}_k$ converge to
\begin{align} \label{eq PropLarge2 1}
    {r}_k\left[i \right]
    &\mathop  \to \limits^{a.s.}
    \sqrt{\Ps}
    \B{w}_k^T
    \B{g}_{{\tt SR},k}
    {x}_k\left[i\right]
    +
    \sqrt{\Ps}
    \sum_{j\neq k}^K
    \B{w}_k^T
    \B{g}_{{\tt SR},j}
    {x}_j\left[i\right]\nonumber\\
    &+
    \B{w}_k^T
    \B{n}_{\tt R}\left[i\right], ~ \text{for both
ZF and MRC/MRT,}
    \\
    {y}_{{\tt D},k}\left[i \right]
    &\mathop  \to \limits^{a.s.}
    \left\{%
\begin{array}{l}
     \sqrt{\frac{\ER}{\sum_{j=1}^K \sigma_{{\tt RD}, j}^{-2}}  }
    {x}_k\left[i-d\right]
    +
    {n}_{{\tt D},k}\left[i\right], ~ \text{for ZF,}\\
    \sqrt{\frac{ \sigma_{{\tt RD}, k}^4\ER}{\sum_{j=1}^K \sigma_{{\tt RD}, j}^2}  }
    {x}_k\left[i-d\right]
    +
    {n}_{{\tt D},k}\left[i\right], ~ \text{for MRC/MRT,}
    \\
\end{array}%
\right. \label{eq PropLarge2 1b}
\end{align}
respectively.
\begin{proof}
With ZF processing, the loop interference is given by
\begin{align} \label{eq ProofLare2 `}
    \sqrt{\PR}
    \B{W}^T
    \B{G}_{\tt RR}
    \B{s}\left[i\right]
    &=
    \sqrt{
        \frac{
            \left(\Nt-K\right)\ER
            }{
            \Nt\sum_{k=1}^K\sigma_{{\tt RD},k}^{-2}
            }
    }
    \B{W}_{\tt ZF}^T
    \frac{
    \B{G}_{\tt RR}
    \hat{\B{G}}_{\tt RD}^{\ast}
        }{
        \Nt
        }    \nonumber
    \\
    &\times\left(\frac{\hat{\B{G}}_{\tt RD}^T \hat{\B{G}}_{\tt RD}^{\ast}}{\Nt}\right)^{-1}
    \B{x}\left[i-d\right]
    \nonumber
    \\
    &\mathop  \to \limits^{a.s.}
    0, ~ \text{as} ~ \Nt\to\infty,
\end{align}
where the convergence follows the law of large numbers. Thus, we
obtain \eqref{eq PropLarge2 1}. By using a similar method as in
Appendix~\ref{Appendix Large1}, we can obtain \eqref{eq PropLarge2
1b}. The results for MRC/MRT processing follow a similar line of
reasoning.
\end{proof}
\end{proposition}

We can see that, by using a very low transmit power, i.e., scaled
proportionally to $1/\Nt$, the loop interference effect at the
receive antennas is negligible [see \eqref{eq PropLarge2 1}].
Although the transmit power is low, the power level of the desired
signal received at each ${\tt D}_k$ is good enough thanks to the
improved array gain, when $\Nt$ grows large.  At the same time,
interpair interference at each ${\tt D}_k$ disappears due to the
orthogonality between the channel vectors [see \eqref{eq
PropLarge2 1b}]. As a result, the quality of the second hop ${\tt
R} \to {\tt D}_k$ is still good enough to provide a robust overall
e2e performance.

\section{Achievable Rate Analysis}

In this section, we derive the e2e achievable rate of the
transmission link ${\tt S}_k \to {\tt R} \to {\tt D}_k$ for ZF and
MRC/MRT processing. The achievable rate is limited by the
weakest/bottleneck link, i.e., it is equal to the minimum of the
achievable rates of the transmissions from ${\tt S}_k$ to ${\tt
R}$ and from ${\tt R}$ to ${\tt D}_k$ \cite{RWW:11:ACSSC}. To
obtain this achievable rate, we use a technique from
\cite{HH:03:IT}. With this technique, the received signal is
rewritten as a known mean gain times the desired symbol, plus an
uncorrelated effective noise whose entropy is upper-bounded by the
entropy of Gaussian noise. This technique is widely used in the
analysis of massive MIMO systems since: i) it yields a simplified
insightful rate expression, which is basically a lower bound of
what can be achieved in practice; and ii) it does not require
instantaneous CSI at the destination
\cite{JAMV:11:WCOM,YM:13:JSAC,AML:12:WCL}. The e2e achievable rate
of the transmission link ${\tt S}_k \to {\tt R} \to {\tt D}_k$ is
given by
\begin{align} \label{eq RateAnalysis 1}
    R_{k}
    =
    \min \left\{
        R_{{\tt SR},k}, R_{{\tt RD},k}
    \right\},
\end{align}
where $R_{{\tt SR},k}$ and $R_{{\tt RD},k}$ are the achievable
rates of the transmission links  ${\tt S}_k \to {\tt R}$ and ${\tt
R} \to {\tt D}_k$, respectively. We next compute $R_{{\tt SR},k}$
and $R_{{\tt RD},k}$. To compute $R_{{\tt SR},k}$,  we consider
\eqref{eq ZFR 1b}. From \eqref{eq ZFR 1b}, the received signal
used for detecting ${x}_k\left[i\right]$ at the relay station can
be written as
\begin{align} \label{eq Rate1a 1}
    {r}_k\left[i \right]
    =
    \underbrace
    {\sqrt{\Ps}
    \E\left\{\B{w}_k^T \B{g}_{{\tt SR},k} \right\}
    {x}_k\left[i\right]}_{\text{desired signal}}
    +
    \underbrace
    {\tilde{{n}}_{{\tt R},k}\left[i\right]}_{\text{effective
    noise}}\!\!,
\end{align}
where $\tilde{{n}}_{{\tt R},k}\left[i\right]$ is considered as the
effective noise, given by
\begin{align} \label{eq en 1}
    &\tilde{{n}}_{{\tt R},k}\left[i\right]
    \triangleq
    \sqrt{\Ps}
    \Big(
    \B{w}_k^T
    \B{g}_{{\tt SR},k}
    -
    \E\left\{\B{w}_k^T \B{g}_{{\tt SR},k} \right\}
    \Big)
    {x}_k\left[i\right]
        \nonumber
    \\
    &+
    \sqrt{\Ps}
    \sum_{j\neq k}^K
    \B{w}_k^T
    \B{g}_{{\tt SR},j}
    {x}_j\left[i\right]
+
    \sqrt{\PR}
    \B{w}_k^T
    \B{G}_{\tt RR}
    \B{s}\left[i\right]
    +
    \B{w}_k^T
    \B{n}_{\tt R}\left[i\right].
\end{align}
We can see that the ``desired signal'' and the ``effective noise''
in \eqref{eq Rate1a 1} are uncorrelated. Therefore, by using the
fact that the worst-case uncorrelated additive noise is
independent Gaussian noise of the same variance, we can obtain an
achievable rate as
\begin{align} \label{eq Rate1 1}
    R_{{\tt SR},k}
    &=
    \log_2
    \left(
        1
        +
        \frac{
            \Ps
            \left|
                \E\left\{\B{w}_k^T \B{g}_{{\tt SR},k} \right\}
            \right|^2
            }{
            \Ps \var\left(\B{w}_k^T \B{g}_{{\tt SR},k}\right)
            +
            {\tt MP}_k
            +
            {\tt LI}_k
            +
            {\tt AN}_k
            }
    \right),
\end{align}
where ${\tt MP}_k$, ${\tt LI}_k$, and ${\tt AN}_k$ represent the
multipair interference, LI, and additive noise effects,
respectively, given by
\begin{align}
    {\tt MP}_k
    &\triangleq
    \Ps
    \sum_{j\neq k}^K
    \E
    \left\{
        \left|\B{w}_k^T
        \B{g}_{{\tt SR},j}\right|^2
    \right\}, \label{eq Rate1 2a}
    \\
    {\tt LI}_k
    &\triangleq
    \PR
    \E
    \left\{
        \left\|
        \B{w}_k^T
        \B{G}_{\tt RR}
        \B{A}
        \right\|^2
    \right\},  \label{eq Rate1 2b}
    \\
    {\tt AN}_k
    &\triangleq
    \E
    \left\{
        \left\|
        \B{w}_k
        \right\|^2
    \right\}.  \label{eq Rate1 2c}
\end{align}
To compute $R_{{\tt RD},k}$, we consider \eqref{eq ZFR 3}.
Following a similar method as in the derivation of $R_{{\tt
SR},k}$,  we obtain
\begin{align} \label{eq LPRate1 2}
    R_{{\tt RD},k}
    \!=\!
    \log_2\!\!
    \left(\!\!
        1
        \!+\!
        \frac{
            \PR
            \left|
                \E\left\{\B{g}_{{\tt RD},k}^T\B{a}_k \right\}
            \right|^2
            }{
            \PR \!\var\left(\!\B{g}_{{\tt RD},k}^T\B{a}_k\!\right)
            \!+\!
            \PR\!\sum\limits_{j\neq k}^K\!\!
            \E\!\!\left\{\!\left|
                \B{g}_{{\tt RD},k}^T\B{a}_j
            \right|^2\!\right\}
            \!+\!
            1
            }
    \!\!\right).
\end{align}

\begin{remark}
The achievable rates in \eqref{eq Rate1 1} and \eqref{eq LPRate1
2} are obtained by approximating  the effective noise  via an
additive Gaussian noise. Since the effective noise is a sum of
many terms, the central limit theorem guarantees that this is a
good approximation, especially in massive MIMO systems. Hence the
rate bounds in \eqref{eq Rate1 1} and \eqref{eq LPRate1 2} are
expected to be quite tight in practice.
\end{remark}

\begin{remark}
The achievable rate \eqref{eq LPRate1 2} is obtained by assuming
that the destination, ${\tt D}_k$ uses only statistical knowledge
of the channel gains $\Big($i.e., $\E\left\{\B{g}_{{\tt
RD},k}^T\B{a}_k \right\}$$\Big)$ to decode the transmitted signals
and, hence, no time, frequency, and power resources need to be
allocated to the transmission of pilots for CSI acquisition.
However, an interesting question is: are our achievable rate
expressions accurate predictors of the system performance? To
answer this question, we compare our achievable rate \eqref{eq
LPRate1 2} with the  ergodic achievable rate of the genie
receiver, i.e., the relay station knows $\B{w}_k^T \B{g}_{{\tt
SR},j}$ and $\B{G}_{\tt RR}$, and the destination ${\tt D}_k$
knows perfectly $\B{g}_{{\tt RD},k}^T \B{a}_j$, $j =1, ..., K$.
For this case, the ergodic e2e achievable rate of the transmission
link ${\tt S}_k \to {\tt R} \to {\tt D}_k$ is
\begin{align} \label{eq upper 1}
    \tilde{R}_{k}
    =
    \min \left\{
        \tilde{R}_{{\tt SR},k}, \tilde{R}_{{\tt RD},k}
    \right\},
\end{align}
where $\tilde{R}_{{\tt SR},k}$ and $\tilde{R}_{{\tt RD},k}$ are
given by
\begin{align} \label{eq upper 3}
    &\tilde{R}_{{\tt SR},k}
    \nonumber\\
    &\!=\!
    \E\left\{\!\!
    \log_2\!\!
    \left(\!\!
        1
        \!+\!
        \frac{
             \Ps \left|\B{w}_k^T \B{g}_{{\tt SR},k}\right|^2
            }{
            \Ps \!\sum\limits_{j\neq k}^K\!\!
                \left|\B{w}_k^T \B{g}_{{\tt SR},j}\right|^2
            \!\!+\!
            \PR\!\left\|\B{w}_k^T\B{G}_{\tt RR}\B{A} \right\|^2
            \!\!+\!
            \left\|\B{w}_k \right\|^2
            }
    \!\!\right)
    \!\!\right\},
\\
    &\tilde{R}_{{\tt RD},k}
    =
    \E\left\{
    \log_2
    \left(
        1
        +
        \frac{
             \PR \left|\B{g}_{{\tt RD},k}^T\B{a}_k\right|^2
            }{
            \PR \sum_{j\neq k}^K
                \left|\B{g}_{{\tt RD},k}^T\B{a}_j\right|^2
            +
            1
            }
    \right)
    \right\}.
\end{align}
In Section~\ref{Sec:Numerical results}, it is demonstrated via
simulations that the performance gap between the achievable rates
given by \eqref{eq RateAnalysis 1} and \eqref{eq upper 1} is
rather small, especially for large $\Nr$ and $\Nt$. Note that the
above ergodic achievable rate in \eqref{eq upper 1} is obtained
under the assumption of perfect CSI which is idealistic in
practice.
\end{remark}

We next provide a new approximate closed-form expression for the
e2e achievable rate given by \eqref{eq RateAnalysis 1} for ZF, and
a new exact one for MRC/MRT processing:

\begin{theorem}\label{Theorem 1}
With ZF processing, the e2e achievable rate of the transmission
link ${\tt S}_k \to {\tt R} \to {\tt D}_k$, for a finite number of
receive antennas at the relay station and $\Nt \gg 1$, can be
approximated as
\begin{align} \label{eq Rate2 re}
    &R_k
    \approx
    R_k^{\tt ZF}\nonumber\\
    &\!\triangleq\!
    \log_2\!
    \left(\!\!
        1
        \!+\!
        \min\left(
        \frac{
            \Ps
            \left(\Nr-K \right) \sigma_{{\tt SR},k}^2
            }{
            \Ps\! \sum\limits_{j=1}^K\!\! \left(\!\beta_{{\tt SR},j}\!-\! \sigma_{{\tt SR},j}^2 \!\right)
            \!+\!
            \PR\sigma_{\tt LI}^2\left(\!1\!-\! K/\Nt \!\right)
            \!+\!
            1
            },
            \right.
            \right.
        \nonumber
        \\
        &\hspace{1.8cm}\left.
        \left.
        \frac{
            \Nt - K
            }{
            \sum_{j=1}^K \sigma_{{\tt RD}, j}^{-2}
            }
        \frac{
            \PR
            }{
            \PR\left(\beta_{{\tt RD},k}- \sigma_{{\tt RD},k}^2 \right) + 1
            }
        \right)
    \right).
\end{align}
\begin{proof}
See Appendix~\ref{sec app thrm1}.
\end{proof}
\end{theorem}

Note that, the above approximation is due to the approximation of
the loop interference. More specifically, to compute the loop
interference term, ${\tt LI}_k$, we approximate $\hat{\B{G}}_{\tt
RD}^T \hat{\B{G}}_{\tt RD}^{\ast}$ as $ \Nt \hat{\B{D}}_{\tt RD}$.
This approximation follows the law of large numbers, and, hence,
becomes exact in the large-antenna limit. In fact, in
Section~\ref{Sec:Numerical results}, we will show that this
approximation is rather tight even for finite number of antennas.

\begin{theorem}\label{Theorem 2}
With MRC/MRT processing, the e2e achievable rate of the
transmission link ${\tt S}_k \to {\tt R} \to {\tt D}_k$, for a
finite number of antennas at the relay station, is given by
\begin{align} \label{eq RateMRC 1}
    R_k
    =
    R_k^{\tt MR}
    &\triangleq
    \log_2
    \left(
        1
        +
        \min\left(
        \frac{
            \Ps
            \Nr \sigma_{{\tt SR},k}^2
            }{
            \Ps \sum_{j=1}^K \beta_{{\tt SR},j}
            +
            \PR\sigma_{\tt LI}^2
            +
            1
            }
        ,\right.\right.\nonumber\\
        &\hspace{2cm}\left.\left.
        \frac{\sigma_{{\tt RD},k}^4}{\sum_{j=1}^K \sigma_{{\tt
        RD},j}^2}
        \frac{
            \PR\Nt
            }{
            \PR\beta_{{\tt RD},k} + 1
            }
        \right)
    \right).
\end{align}
\begin{proof}
See Appendix~\ref{sec app thrm2}.
\end{proof}
\end{theorem}

\section{Performance Evaluation}

To evaluate the system performance, we consider the sum spectral
efficiency. The sum spectral efficiency is defined as the sum-rate
(in bits) per channel use. Let $T$ be the length of the coherence
interval (in symbols). During each coherence interval, we spend
$\tau$ symbols for training, and the remaining interval is used
for the payload data transmission. Therefore, the sum spectral
efficiency is given by
\begin{align} \label{eq spect 1}
    \mathcal{S}_{\mathrm{FD}}^{\tt A}
    &\triangleq
    \frac{T-\tau}{T}
    \sum_{k=1}^K R_k^{\tt A},
\end{align}
where ${\tt A} \in \left\{{\tt ZF}, {\tt MR}\right\}$ corresponds
to ZF and MRC/MRT processing.  Note that in the case of ZF
processing, $R_k^{\tt ZF}$ is an approximate result. However, in
the numerical results (see Section~\ref{Sec:Numerical results 1}),
we show that this approximation is very tight and fairly accurate.
For this reason, and without significant lack of clarity, we
hereafter consider the rate results of ZF
processing as exact. 

From \emph{Theorems}~\ref{Theorem 1}, \ref{Theorem 2}, and
\eqref{eq spect 1}, the sum spectral efficiencies of ZF and
MRC/MRT processing for the full-duplex mode are, respectively,
given by \eqref{eq spectFD 1a} and \eqref{eq spectFD 1ab} shown at
the top of the next page.

\setcounter{eqnback}{\value{equation}} \setcounter{equation}{37}
\begin{figure*}[!t]
\begin{align} \label{eq spectFD 1a}
    \mathcal{S}_{\mathrm{FD}}^{\tt ZF}
    &=
    \frac{T-\tau}{T}\!
    \sum_{k=1}^K\!
    \log_2\!\!
    \left(\!
        1
        +
        \min\left(\!
        \frac{
            \Ps
            \left(\Nr-K \right) \sigma_{{\tt SR},k}^2
            }{
            \Ps \!\sum_{j=1}^K \!\left(\beta_{{\tt SR},j}\!-\! \sigma_{{\tt SR},j}^2 \right)
            \!+\!
            \PR\!\sigma_{\tt LI}^2\left(1\!- \!K/\Nt \right)
            \!+\!
            1
            },
        \frac{
            \Nt - K
            }{
            \sum_{j=1}^K \sigma_{{\tt RD}, j}^{-2}
            }
        \frac{
            \PR
            }{
            \PR\!\left(\!\beta_{{\tt RD},k}\!-\! \sigma_{{\tt RD},k}^2 \!\right) \!+ \!1
            }
        \!\right)
    \!\!\right),
    \\ \label{eq spectFD 1ab}
    \mathcal{S}_{\mathrm{FD}}^{\tt MR}
    &=
    \frac{T-\tau}{T}
    \sum_{k=1}^K
    \log_2
    \left(\!
        1
        \!+\!
        \min\left(
        \frac{
            \Ps
            \Nr \sigma_{{\tt SR},k}^2
            }{
            \Ps \sum_{j=1}^K \beta_{{\tt SR},j}
            +
            \PR\sigma_{\tt LI}^2
            +
            1
            }
        ,
        \frac{\sigma_{{\tt RD},k}^4}{\sum_{j=1}^K \sigma_{{\tt
        RD},j}^2}
        \frac{
            \PR\Nt
            }{
            \PR\beta_{{\tt RD},k} + 1
            }
        \right)\!
    \right).
\end{align}
\hrulefill
\end{figure*}
\setcounter{eqncnt}{\value{equation}}
\setcounter{equation}{\value{eqnback}}


\subsection{Power Efficiency}
In this part, we study the potential for power savings by using
very large antenna arrays at the relay station.
\begin{enumerate}

\item \emph{Case I}: We consider the case where $\Pp$ is fixed,
$\Ps = \ES/\Nr$, and $\PR=\ER/\Nt$, where $\ES$ and
    $\ER$ are fixed regardless of $\Nr$ and $\Nt$. This case corresponds to the case where the channel estimation accuracy is fixed, and we want to investigate the potential for power saving
    in the data transmission phase. When $\Nt$ and
    $\Nr$ go to infinity with the same speed, the sum spectral efficiencies of ZF and MRC/MRT processing can be expressed as
\setcounter{equation}{39}\begin{align} \label{eq spectFD as5}
    \mathcal{S}_{\mathrm{FD}}^{\tt ZF}
    &\to
    \frac{T\!-\!\tau}{T}\!
    \sum_{k=1}^K\!
    \log_2\!\!
    \left(\!
        1
        \!+\!
        \min\left(\!
            \ES \sigma_{{\tt SR},k}^2
            ,
        \frac{
            \ER
            }{
            \sum_{j=1}^K\!\! \sigma_{{\tt RD}, j}^{-2}
            }
        \!\right)\!\!
    \right)\!\!,
    \\ \label{eq spectFD as6}
    \mathcal{S}_{\mathrm{FD}}^{\tt MR}
    &\to
    \frac{T\!-\!\tau}{T}\!\!
    \sum_{k=1}^K\!
    \log_2\!\!
    \left(\!
        1
        \!+\!
        \min\left(\!
            \ES
            \sigma_{{\tt SR},k}^2
        ,
        \frac{\sigma_{{\tt RD},k}^4\ER}{\sum_{j=1}^K \!\sigma_{{\tt
        RD},j}^2}
        \!\right)
    \!\!\right)\!\!.
\end{align}

The expressions in \eqref{eq spectFD as5} and \eqref{eq spectFD
as6} show that, with large antenna arrays,  we can reduce the
transmitted power of each source and of the relay station
proportionally to $1/\Nr$ and $1/\Nt$, respectively, while
maintaining a given QoS. If we now assume that large-scale fading
is neglected (i.e., $\beta_{{\tt SR},k}= \beta_{{\tt RD},k} =1,
\forall k$), then from \eqref{eq spectFD as5} and \eqref{eq
spectFD as6}, the asymptotic performances of ZF and MRC/MRT
processing are the same and given by:
\begin{align} \label{eq spectFD as7}
    \mathcal{S}_{\mathrm{FD}}^{\tt A}
    &\to
    \frac{T-\tau}{T}
    K
    \log_2
    \left(\!
        1
        +
        \sigma_1^2
        \min\left(\!
            \ES
            ,
        \frac{
            \ER
            }{
            K
            }
        \!\right)
    \!\!\right),
\end{align}
where $\sigma_1^2 \triangleq \frac{\tau\Pp}{\tau\Pp+1}$. The sum
spectral efficiency in \eqref{eq spectFD as7} is equal to the one
of $K$ parallel single-input single-output channels with transmit
power $ \sigma_1^2
        \min\left(
            \ES
            ,
        \frac{
            \ER
            }{
            K
            }\right)$, without interference and  fast
            fading. We see that, by using large antenna
            arrays, not only the transmit powers are reduced
            significantly, but also the sum spectral efficiency is increased $K$ times (since
            all $K$ different
communication pairs are served simultaneously).

\item \emph{Case II}: If $\Pp=\Ps = \ES/\sqrt{\Nr}$ and
$\PR=\ER/\sqrt{\Nt}$, where $\ES$ and
    $\ER$ are fixed regardless of $\Nr$ and $\Nt$. When
    $\Nr$ goes to infinity and $\Nt = \kappa \Nr$, the sum spectral
    efficiencies
    converge to
\begin{align} \label{eq spectFD as8}
    &\mathcal{S}_{\mathrm{FD}}^{\tt ZF}
    \nonumber\\
    &\to
    \frac{T\!-\!\tau}{T}\!\!
    \sum_{k=1}^K\!
    \log_2\!\!
    \left(\!
        1
        \!+\!
        \min\left(\!
            \tau\ES^2\beta_{{\tt SR},k}^2
            ,
        \frac{
            \sqrt{\kappa} \tau\ES\ER
            }{
            \sum_{j=1}^K \beta_{{\tt RD}, j}^{-2}
            }
        \!\right)
    \!\!\right),
    \\ \label{eq spectFD as9}
    &\mathcal{S}_{\mathrm{FD}}^{\tt MR}
    \nonumber\\
    &\to
     \frac{T\!-\!\tau}{T}\!\!
    \sum_{k=1}^K\!\!
    \log_2\!\!
    \left(\!\!
        1
        \!+\!
        \min\!\!\left(\!\!
            \tau\ES^2\beta_{{\tt SR},k}^2
            ,
        \frac{
            \sqrt{\kappa} \tau\ES\ER \beta_{{\tt RD},k}^4
            }{
            \sum_{j=1}^K \beta_{{\tt RD}, j}^{-2}
            }
        \!\right)\!\!
    \right)\!.
\end{align}
We see that, if the transmit powers of the uplink training and
data transmission are the same, (i.e., $\Pp=\Ps$), we cannot
reduce  the transmit powers of each source and of the relay
station as aggressively as in \emph{Case I} where the pilot power
is kept fixed. Instead, we can scale down the transmit powers of
each source and of the relay station proportionally to only
$1/\sqrt{\Nr}$ and $1/\sqrt{\Nt}$, respectively. This observation
can be interpreted as, when we cut the transmitted power of each
source, both the data signal and the pilot signal suffer from
power reduction, which leads to the so-called ``squaring effect"
on the spectral efficiency \cite{HH:03:IT}.
\end{enumerate}

\subsection{Comparison between Half-Duplex and Full-Duplex Modes}
In this section, we compare the performance of the half-duplex and
full-duplex modes. For the half-duplex mode, two orthogonal time
slots are allocated for two transmissions: sources to the relay
station and the relay station to destinations \cite{SNDYL:13:ICC}.
The half-duplex mode does not induce the loop interference at the
cost of imposing a pre-log factor $1/2$ on the spectral
efficiency. The sum spectral efficiency of the half-duplex mode
can be obtained directly from \eqref{eq spectFD 1a} and \eqref{eq
spectFD 1ab} by neglecting the loop interference effect. Note
that, with the half-duplex mode, the sources and the relay station
transmit only half of the time compared to the full-duplex mode.
For fair comparison, the total energies spent in a coherence
interval for both modes are set to be the same. As a result, the
transmit powers of each source and of the relay station used in
the half-duplex mode are double the powers used in the full-duplex
mode and, hence, the sum spectral efficiencies of the half-duplex
mode for ZF and MRC/MRT processing are respectively given
by\footnote{
    Here, we assume that the relay station in the half-duplex mode employs the same number of transmit and receive antennas as in the full-duplex mode.
    This assumption corresponds to the ``RF chains conserved" condition, where an equal number of total RF chains
are assumed \cite[Section III]{AKSRC:12:MobiCom}. Note that, in
order to receive the transmitted signals from the destinations
during the channel estimation phase, additional ``receive RF
chains'' have to be used in the transmit array for both
full-duplex and half-duplex cases. The comparison between
half-duplex and full-duplex modes can be also performed with the
``number of antennas preserved'' condition, where the number of
antennas at the relay station used in the half-duplex mode is
equal to the total number of transmit and receive antennas used in
the FD mode, i.e., is equal to $\Nt+\Nr$. However, the cost of the
required RF chains is significant as opposed to adding an extra
antenna. Thus, we choose the ``RF chains conserved'' condition for
our comparison.}
\begin{align} \label{eq spectHD 1a}
    \mathcal{S}^{\tt ZF}_{\mathrm{HD}}
    &\!=\!
    \frac{T\!-\!\tau}{2T}\!
    \sum_{k=1}^K\!
    \log_2\!\!
    \left(\!\!
        1
        \!+\!
        \min\!\left(\!
        \frac{
            2\Ps
            \left(\Nr-K \right) \sigma_{{\tt SR},k}^2
            }{
            2\Ps \sum_{j=1}^K\!\! \left(\beta_{{\tt SR},j}\!-\! \sigma_{{\tt SR},j}^2 \right)
            +
            1
            },\right.\right.\nonumber\\ &\left.\left.
        \frac{
            \Nt - K
            }{
            \sum_{j=1}^K \!\sigma_{{\tt RD}, j}^{-2}
            }
        \frac{
            2\PR
            }{
            2\PR\!\left(\!\beta_{{\tt RD},k}\!- \!\sigma_{{\tt RD},k}^2 \!\right) + 1
            }
        \!\!\right)\!\!
    \!\right)\!\!,
    \\ \label{eq spectFD 1b}
    \mathcal{S}^{\tt MR}_{\mathrm{HD}}
    &=
    \frac{T-\tau}{2T}
    \sum_{k=1}^K
    \log_2
    \left(
        1
        +
        \min\left(
        \frac{
            2\Ps
            \Nr \sigma_{{\tt SR},k}^2
            }{
            2\Ps \sum_{j=1}^K \beta_{{\tt SR},j}
            +
            1
            }
        ,\right.\right.\nonumber\\ &\left.\left.
        \frac{\sigma_{{\tt RD},k}^4}{\sum_{j=1}^K \sigma_{{\tt
        RD},j}^2}
        \frac{
            2\PR \Nt
            }{
            2\PR\beta_{{\tt RD},k} + 1
            }
        \right)
    \right).
\end{align}
Depending on the transmit powers, channel gains, channel
estimation accuracy, and the loop interference level, the
full-duplex mode is preferred over the half-duplex modes and vice
versa. The critical factor is the loop interference level. If all
other factors are fixed, the full-duplex mode outperforms the
half-duplex mode if $\sigma_{\tt LI}^2 \leq \sigma_{\tt LI, 0}^2$,
where $\sigma_{\tt LI, 0}^2$ is the root of $\mathcal{S}^{\tt
ZF}_{\mathrm{FD}} = \mathcal{S}^{\tt ZF}_{\mathrm{HD}}$ for the ZF
processing or the root of $\mathcal{S}^{\tt MR}_{\mathrm{FD}} =
\mathcal{S}^{\tt MR}_{\mathrm{HD}}$ for the MRC/MRT processing.

From the above observation, we propose to use a hybrid relaying
mode as follows:
$$
{\tt Hybrid ~ Relaying ~ Mode} \!=\!
\left\{\!\!%
\begin{array}{l}
  {\tt Full-Duplex, ~ if} ~ \mathcal{S}^{\tt A}_{\mathrm{FD}} \!\geq\! \mathcal{S}^{\tt
A}_{\mathrm{HD}}\\
  {\tt Half-Duplex, ~otherwise}. \\
\end{array}%
\right.
$$
Note that, with hybrid relaying, the relaying mode is chosen for
each large-scale fading realization.

\vspace{-0.cm}

\subsection{Power Allocation}\label{sec: PA}\vspace{-0.2cm}
In previous sections, we assumed that the transmit powers of all
users are the same. The system performance can be improved by
optimally allocating different powers to different sources. Thus,
in this section, we assume that the transmit powers of different
sources are different. We assume that the design for training
phase is done in advance, i.e., the training duration, $\tau$, and
the pilot power, $\Pp$, were determined. We are interested in
designing a power allocation algorithm in the data transmission
phase that maximizes the energy efficiency, subject to a given sum
spectral efficiency and the constraints of maximum powers
transmitted from sources and the relay station, for each
large-scale realization. The energy efficiency (in bits/Joule) is
defined as the sum spectral efficiency divided by the total
transmit power. Let the transmit power of the $k$th source be
$p_{{\tt S},k}$. Therefore, the energy efficiency of the
full-duplex mode is given by
\begin{align}\label{eq: EE1}
    {\tt EE}^{\tt A}
    \triangleq
    \frac{\mathcal{S}^{\tt A}_{\mathrm{FD}}}
    {\frac{T-\tau}{T}\left(\sum_{k=1}^K p_{{\tt S},k} +
    \PR\right)}.
\end{align}
Mathematically, the optimization problem can be formulated as
\begin{align}\label{eq opt 1}
    \left.%
\begin{array}{l}
  \text{maximize} \hspace{1 cm} {\tt EE}^{\tt A}  \\
  \text{subject to}
     \hspace{1 cm} \mathcal{S}^{\tt A}_{\mathrm{FD}} = \mathcal{S}^{\tt A}_0\\
     \hspace{2.4 cm} 0\leq\Psk \leq p_0, k=1, ..., K \\
     \hspace{2.4 cm} 0\leq\PR \leq p_1\\
\end{array}%
\right.
\end{align}
where $\mathcal{S}^{\tt A}_0$ is a required sum spectral
efficiency, while $p_0$ and $p_1$ are the peak power constraints
of $p_{{\tt S},k}$ and $\PR$, respectively.

From \eqref{eq spectFD 1a}, \eqref{eq spectFD 1ab}, and \eqref{eq:
EE1}, the optimal power allocation problem in \eqref{eq opt 1} can
be rewritten as
\begin{align}\label{eq opt 2}
    \left.%
\begin{array}{l}
  \text{minimize} \hspace{1 cm} \sum_{k=1}^K p_{{\tt S},k} + \PR  \\
  \text{subject to}\\
        \hspace{0.8 cm} \frac{T\!-\!\tau}{T}\!\!\sum\limits_{k=1}^K\!\log_2\!\!\left(\!\!1\!+\!\min \left\{\frac{a_k\Psk}{\sum\limits_{j=1}^K b_j p_{{\tt S},j} + c_k\PR +1 }, \frac{d_k\PR}{e_k\PR + 1}
        \right\}\!\right)\!=\! \mathcal{S}^{\tt A}_0\\
  \hspace{0.8 cm} 0\leq\Psk \leq p_0, k=1, ..., K \\
     \hspace{0.8 cm} 0\leq\PR \leq p_1\\
\end{array}%
\right.
\end{align}
where $a_k$, $b_k$, $c_k$, $d_k$, and $e_k$ are constant values
(independent of the transmit powers) which are different for ZF
and MRC/MRT processing. More precisely,
\begin{itemize}
    \item For ZF: $a_k= \left(\Nr-K \right)\sigma_{{\tt SR},k}^2$,
    $b_k = \beta_{{\tt SR},k}- \sigma_{{\tt SR},k}^2$, $c_k=\sigma_{\tt LI}^2\left(1- K/\Nt
    \right)$, $d_k=         \frac{
            \Nt - K
            }{
            \sum_{j=1}^K \sigma_{{\tt RD}, j}^{-2}
            }$, and $e_k= \beta_{{\tt RD},k}- \sigma_{{\tt RD},k}^2$.
    \item For MRC/MRT: $a_k=\Nr \sigma_{{\tt SR},k}^2$, $b_k =
    \beta_{{\tt SR},k}$, $c_k=\sigma_{\tt LI}^2$, $d_k =         \frac{\sigma_{{\tt RD},k}^4}{\sum_{j=1}^K \sigma_{{\tt
        RD},j}^2} \Nt$, and $e_k=\beta_{{\tt RD},k}$.
\end{itemize}

\begin{figure}[t]
    \centerline{\includegraphics[width=0.48\textwidth]{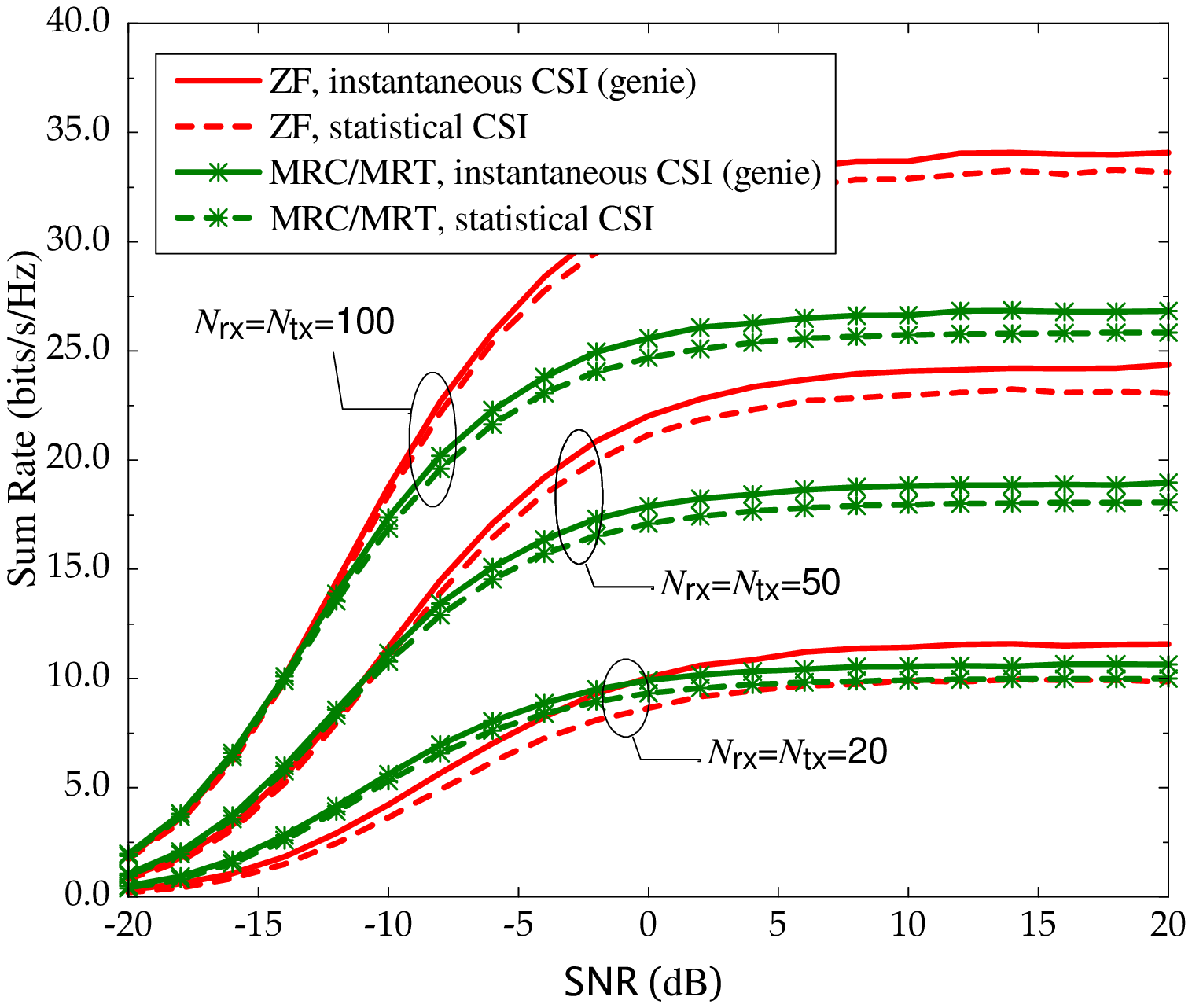}}
    \caption{Sum rate versus $\mathsf{SNR}$ for ZF and MRC/MRT processing ($K=10$, $\tau=2K$, and $\sigma_{\tt LI}^2=1$).}
    \label{fig:2}
\end{figure}

The problem \eqref{eq opt 2} is equivalent to
\begin{align}\label{eq opt 4}
    \left.%
\begin{array}{l}
  \text{minimize} \hspace{0.5 cm} \sum_{k=1}^K p_{{\tt S},k} + \PR   \\
  \text{subject to}
        \hspace{0.5 cm}
        \frac{T-\tau}{T}\sum_{k=1}^K\log_2\left(1+\gamma_k\right)= \mathcal{S}^{\tt A}_0\\
        \hspace{2.2 cm} \gamma_k \leq \frac{a_k\Psk}{\sum_{j=1}^K b_j p_{{\tt S},j} + c_k\PR +1},  k=1, ..., K\\
      \hspace{2.2 cm}  \gamma_k \leq \frac{d_k\PR}{e_k\PR + 1},  k=1, ..., K\\
      \hspace{2.25 cm} 0\leq\Psk \leq p_0, k=1, ..., K\\
     \hspace{2.25 cm} 0\leq\PR \leq p_1.\\
\end{array}%
\right.
\end{align}
Since $a_k$, $b_k$, $c_k$, $d_k$, and $e_k$ are positive,
\eqref{eq opt 4} can be equivalently written as
\begin{align}\label{eq opt 5}
    \left.%
\begin{array}{l}
  \text{minimize} \hspace{0.5 cm} \sum_{k=1}^K p_{{\tt S},k} + \PR   \\
  \text{subject to}
      \hspace{0.5 cm} \prod_{k=1}^K\left(1+\gamma_k\right) = 2^{\frac{T \mathcal{S}^{\tt A}_0}{T-\tau}}\\
      \hspace{0.5 cm}  \sum\limits_{j=1}^K \!\!\frac{b_j}{a_k} p_{{\tt S},j}\gamma_k \Psk^{-1} \!+\! \frac{c_k}{a_k}\PR\gamma_k\Psk^{-1}+ \frac{1}{a_k}\gamma_k\Psk^{-1} \leq 1,  \forall k\\
      \hspace{0.5 cm}  \frac{e_k}{d_k}\gamma_k + \frac{1}{d_k}\gamma_k\PR^{-1} \leq 1,  k=1, ..., K\\
      \hspace{0.5 cm} 0\leq\Psk \leq p_0, k=1, ..., K,\\
      \hspace{0.5 cm} 0\leq\PR \leq p_1.\\
\end{array}%
\right.
\end{align}
We can see that the objective function and the inequality
constraints are posynomial functions. If the equality constraint
is a monomial function, the problem \eqref{eq opt 5} becomes a GP
which can be reformulated as a convex problem, and can be solved
efficiently by using convex optimization tools, such as CVX
\cite{BV:04:Book}. However, the equality constraint in \eqref{eq
opt 5} is a posynomial function, so we cannot solve \eqref{eq opt
5} directly using convex optimization tools. Yet, by using the
technique in \cite{WCLE:11:VT}, we can efficiently find an
approximate solution of \eqref{eq opt 5} by
 solving a sequence of GPs. More precisely, from
 \cite[Lemma~1]{WCLE:11:VT}, we can use $\kappa_k\gamma_k^{\eta_k}$ to approximate
 $1+\gamma_k$ near a point $\hat{\gamma}_k$, where $\eta_k \triangleq \hat{\gamma}_k\left(1+\hat{\gamma}_k
 \right)^{-1}$ and $\kappa_k \triangleq \hat{\gamma}_k^{-\eta_k}\left(1+ \hat{\gamma}_k
 \right)$. As a consequence, near a point $\hat{\gamma}_k$, the left hand side of the equality constraint can be
 approximated as
\begin{align}\label{eq opt 6}
    \prod_{k=1}^K\left(1+\gamma_k\right)
    \approx
    \prod_{k=1}^K
    \kappa_k\gamma_k^{\eta_k},
\end{align}
which is a monomial function. Thus, by using the local
approximation given by \eqref{eq opt 6}, the optimization problem
\eqref{eq opt 5} can be approximated by a GP. By using a similar
technique as in \cite{WCLE:11:VT}, we formulate the following
algorithm to solve \eqref{eq opt 5}:

\hrulefill 
\begin{algorithm}[Successive approximation algorithm for \eqref{eq opt
5}]\label{sec: algo} ~ \vspace{-0.5cm}

\hrulefill
\begin{description}
  \item[1.] \emph{Initialization}: set $i=1$, choose the initial values of
  $\gamma_k$ as $\gamma_{k,1}$, $k=1, ..., K$. Define a tolerance $\epsilon$, the maximum number of iterations $L$, and parameter $\alpha$.
  \item[2.] \emph{Iteration} $i$: compute $\eta_{k,i} = {\gamma}_{k,i}\left(1+{\gamma}_{k,i}
 \right)^{-1}$ and $\kappa_{k,i} = {\gamma}_{k,i}^{-\eta_{k,i}}\left(1+
 {\gamma}_{k,i}
 \right)$. Then, solve the GP:
\begin{align*}
    \left.%
\begin{array}{l}
  \text{minimize} \hspace{0.5 cm} \sum_{k=1}^K p_{{\tt S},k} + \PR   \\
  \text{subject to}
      \hspace{0.5 cm}     \prod_{k=1}^K \kappa_{k,i}\gamma_k^{\eta_{k,i}} = 2^{\frac{T \mathcal{S}^{\tt A}_0}{T-\tau}}\\
      \hspace{0.2 cm}  \sum\limits_{j=1}^K \!\!\frac{b_j}{a_k} p_{{\tt S},j}\gamma_k \Psk^{-1} \!+\! \frac{c_k}{a_k}\PR\gamma_k\Psk^{-1}\!+\! \frac{1}{a_k}\gamma_k\Psk^{-1} \leq 1, \forall k\\
      \hspace{0.2 cm}  \frac{e_k}{d_k}\gamma_k + \frac{1}{d_k}\gamma_k\PR^{-1} \leq 1,  k=1, ..., K\\
      \hspace{0.2 cm} 0\leq\Psk \leq p_0, k=1, ..., K, ~ 0\leq\PR \leq p_1\\
     \hspace{0.2 cm} \alpha^{-1}\gamma_{k,i}\leq\gamma_k \leq \alpha\gamma_{k,i}\\
\end{array}%
\right.
\end{align*}
 Let $\gamma_k^{\ast}$, $k=1, ..., K$ be the solutions.
  \item[3.] If $\max_{k}\left|\gamma_{k,i} - \gamma_k^{\ast} \right| < \epsilon$ or $i=L$ $\rightarrow$ Stop. Otherwise, go to step 4.
  \item[4.] Set $i = i+1$, $\gamma_{k,i} = \gamma_k^{\ast}$, go
  to step 2.
\end{description}
\end{algorithm}

\hrulefill

\begin{figure}[t]
    \centerline{\includegraphics[width=0.48\textwidth]{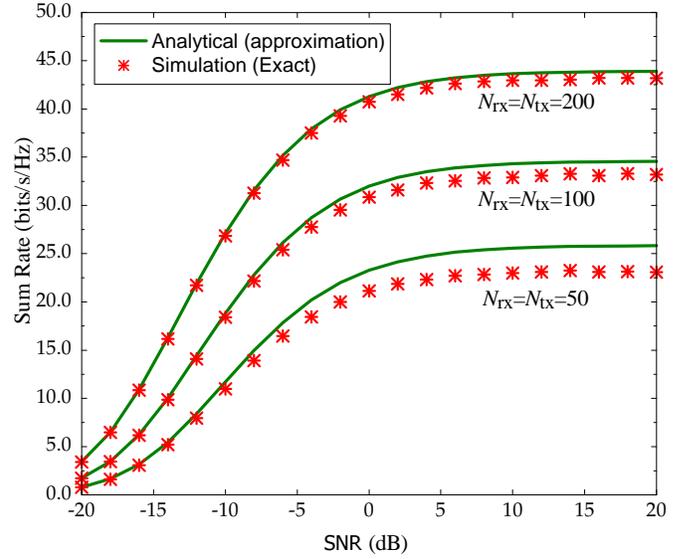}}
    \caption{Sum rate versus $\mathsf{SNR}$ for ZF processing ($K=10$, $\tau=2K$, and $\sigma_{\tt LI}^2=1$).}
    \label{fig:3}
\end{figure}

Note that the parameter $\alpha > 1$ is used to control the
approximation accuracy in \eqref{eq opt 6}. If $\alpha$ is close
to $1$, the accuracy is high, but the convergence speed is low and
vice versa if $\alpha$ is large. As discussed in
\cite{WCLE:11:VT}, $\alpha=1.1$ offers a good accuracy and
convergence speed tradeoff.

\section{Numerical Results} \label{Sec:Numerical results}
In all illustrative examples, we choose the length of the
coherence interval to be $T=200$ (symbols), the number of
communication pairs $K=10$, the training length $\tau = 2K$, and
$\Nt=\Nr$. Furthermore, we define $\mathsf{SNR} \triangleq \Ps$.

\subsection{Validation of Achievable Rate Results}  \label{Sec:Numerical results 1}
In this subsection, we evaluate the validity of our achievable
rate given by \eqref{eq RateAnalysis 1} as well as the
approximation used to derive the closed-form expression given in
Theorem~\ref{Theorem 1}.  We choose the loop interference level
$\sigma_{\tt LI}^2=1$. We assume that $\Pp=\Ps$, and that the
total transmit power of the $K$ sources is equal to the transmit
power of the relay station, i.e., $\PR=K\Ps$.

We first compare our achievable rate given by \eqref{eq
RateAnalysis 1}, where the destination uses the statistical
distributions of the channels (i.e., the means of channel gains)
to detect the transmitted signal, with the one obtained by
\eqref{eq upper 1}, where we assume that there is a genie receiver
(instantaneous CSI) at the destination. Figure~\ref{fig:2} shows
the sum rate versus $\mathsf{SNR}$ for ZF and MRC/MRT processing.
The dashed lines represent the sum rates obtained numerically from
\eqref{eq RateAnalysis 1}, while the solid lines represent the
ergodic sum rates obtained from \eqref{eq upper 1}. We can see
that the relative performance gap between the cases with
instantaneous (genie) and statistical CSI at the destinations is
small. For example, with $\Nr=\Nt=50$, at $\mathsf{SNR}=5$dB, the
sum-rate gaps are 0.65 bits/s/Hz and 0.9 bits/s/Hz for MRC/MRT and
ZF processing, respectively. This implies that using the mean of
the effective channel gain for signal detection is fairly
reasonable, and the achievable rate given in \eqref{eq
RateAnalysis 1} is a good predictor of the system performance.

Next, we evaluate the validity of the approximation given by
\eqref{eq Rate2 re}. Figure~\ref{fig:3} shows the sum rate versus
$\mathsf{SNR}$ for different numbers of transmit (receive)
antennas. The ``Analytical (approximation)'' curves are obtained
by using Theorem~\ref{Theorem 1}, and the ``Simulation (exact)''
curves are generated from the outputs of a Monte-Carlo simulator
using \eqref{eq RateAnalysis 1}, \eqref{eq Rate1 1}, and \eqref{eq
LPRate1 2}. We can see that the proposed approximation is very
tight, especially for large antenna arrays.

\begin{figure}[t]
    \centerline{\includegraphics[width=0.48\textwidth]{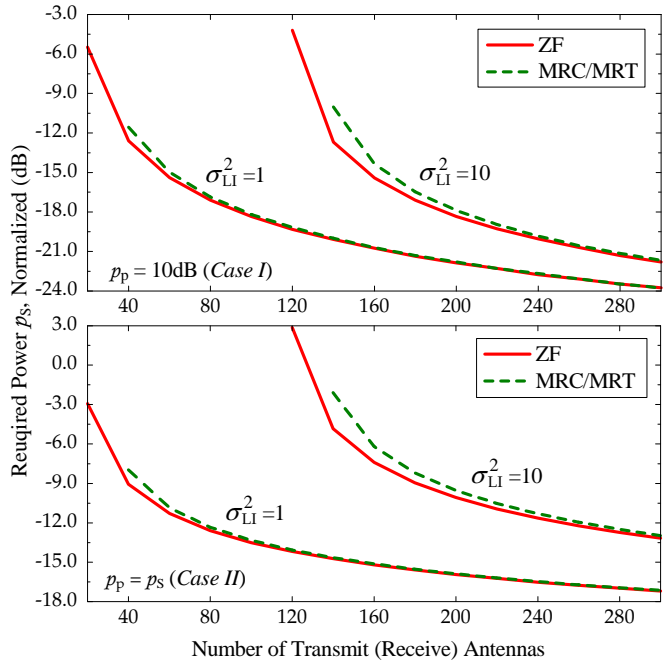}}
    \caption{Transmit power, $\Ps$, required to achieve 1 bit/channel use per user for ZF and MRC/MRT processing ($K=10$, $\tau=2K$, and $\PR=K\Ps$).}
    \label{fig:4}
\end{figure}

\subsection{Power Efficiency}
We now examine the power efficiency of using large antenna arrays
for two cases: $\Pp$ is fixed (\emph{Case I}) and $\Pp = \Ps$
(\emph{Case II}). We will examine how much transmit power is
needed to reach a predetermined sum spectral efficiency. We set
$\PR=K\Ps$ and $\beta_{{\tt SR}, k} = \beta_{{\tt RD}, k} =1$,
$k=1, 2, ..., K$. Figure~\ref{fig:4} shows the required transmit
power, $\Ps$, to achieve $1$ bits/s/Hz per communication pair. We
can see that when the number of antennas increases, the required
transmit powers are significantly reduced. As predicted by the
analysis, in the large-antenna regime, we can cut back the power
by approximately $3$dB and $1.5$dB by doubling the number of
antennas for \emph{Case I} and \emph{Case II}, respectively. When
the loop interference is high and the number of antennas is
moderate, the power efficiency can benefit more by increasing the
number of antennas. For instance, for $\sigma_{\tt LI}^2=10$,
increasing the number of antennas from $120$ to $240$ yields a
power reduction of $15$dB and $13$dB for \emph{Case I} and
\emph{Case II}, respectively.  Regarding the loop interference
effect, when $\sigma_{\tt LI}^2$ increases, we need more transmit
power. However, when $\sigma_{\tt LI}^2$ is high and the number of
antennas is small, even if we use infinite transmit power, we
cannot achieve a required sum spectral efficiency. Instead of
this, we can add more antennas to reduce the loop interference
effect and achieve the required QoS. Furthermore, when the number
of antennas is large, the difference in performance between ZF and
MRC/MRT processing is negligible.

\begin{figure}[t]
    \centerline{\includegraphics[width=0.48\textwidth]{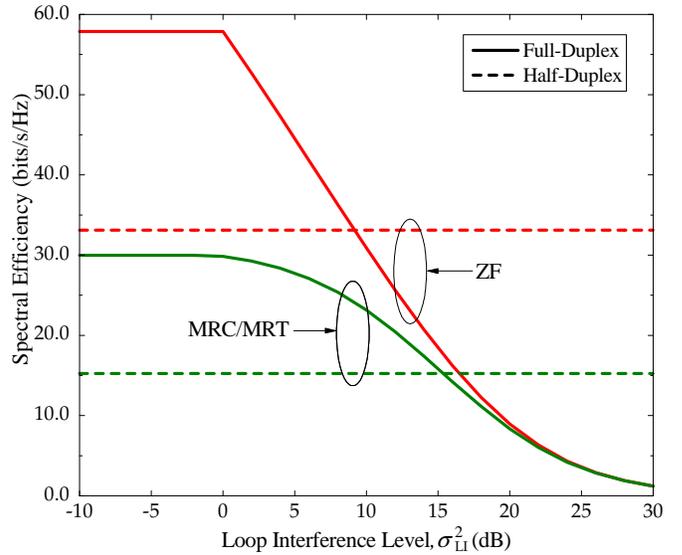}}
    \caption{Sum spectral efficiency versus the loop interference
levels for half-duplex and full-duplex relaying ($K=10$,
$\tau=2K$, $\PR=\Pp=\Ps=10$dB, and $\Nt=\Nr=100$).}
    \label{fig:6}
\end{figure}

\begin{figure}[t]
    \centerline{\includegraphics[width=0.48\textwidth]{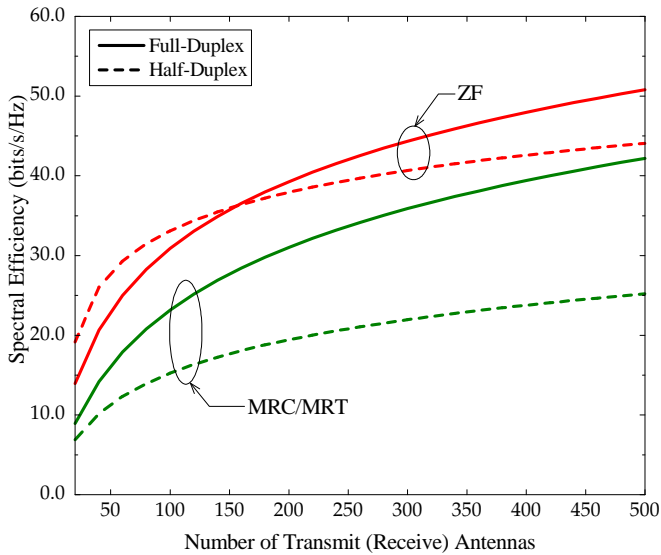}}
    \caption{Sum spectral efficiency versus the number of transmit
(receive) antennas for half-duplex and full-duplex relaying
($K=10$, $\tau=2K$, $\PR=\Pp=\Ps=10$dB, and $\sigma_{\tt
LI}^2=10$dB).}
    \label{fig:7}
\end{figure}

\subsection{Full-Duplex Vs. Half-Duplex, Hybrid Relaying Mode}

\begin{figure}[t]
    \centerline{\includegraphics[width=0.48\textwidth]{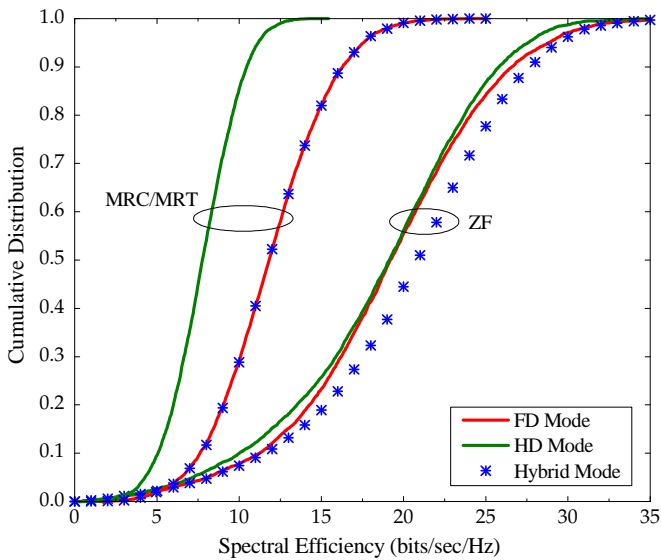}}
    \caption{Cumulative distribution of the sum spectral efficiency for half-duplex, full-duplex, and Hybrid relaying ($K=10$, $\tau=2K$, $\PR=\Pp=\Ps=10$dB, and $\sigma_{\tt LI}^2=10$dB).}
    \label{fig:8}
\end{figure}

Firstly, we compare the performance between half-duplex and
full-duplex relaying for different loop interference levels,
$\sigma_{\tt LI}^2$. We choose $\PR=\Pp=\Ps=10$dB, $\beta_{{\tt
SR}, k} = \beta_{{\tt RD}, k} =1$, $\forall k$, and $\Nr=\Nt=100$.
Figure~\ref{fig:6} shows the sum spectral efficiency versus the
loop interference levels for ZF and MRC/MRT. As expected, at low
$\sigma_{\tt LI}^2$, full-duplex relaying outperforms half-duplex
relaying. This gain is due to the larger pre-log factor (one) of
the full-duplex mode. However, when $\sigma_{\tt LI}^2$ is high,
loop interference dominates the system performance of the
full-duplex mode and, hence, the performance of the half-duplex
mode is superior. In this case, by using larger antenna arrays at
the relay station, we can reduce the effect of the loop
interference and exploit the larger pre-log factor of the
full-duplex mode. This fact
 is illustrated in Fig.~\ref{fig:7} where the sum spectral
efficiency is represented as a function of the number of antennas,
at $\sigma_{\tt LI}^2=10$dB.

We next consider a more practical scenario that incorporates
small-scale fading and large-scale fading. The large-scale fading
is modeled by path loss, shadow fading, and random source and
destination locations. More precisely, the large-scale fading
$\beta_{{\tt SR},k}$ is
\begin{align*}
\beta_{{\tt SR},k}= \frac{z_{{\tt
SR},k}}{1+\left(\ell_k/\ell_0\right)^\nu},
\end{align*}
where $z_{{\tt SR},k}$ represents a log-normal random variable
with standard deviation of $\sigma$dB, $\nu$ is the path loss
exponent, $\ell_k$ denotes the distance between ${\tt S}_k$ and
the receive array of the relay station, and $\ell_0$ is a
reference distance. We use the same channel model for $\beta_{{\tt
RD},k}$.

We assume that all sources and destinations are located uniformly
at random inside a disk with a diameter of $1000$m. For our
simulation, we choose $\sigma =8$dB, $\nu=3.8$, $\ell_0=200$m,
which are typical values in an urban cellular environment
\cite{CA:08:WCOM}. Furthermore, we choose $\Nr=\Nt=200$,
$\PR=\Pp=\Ps=10$dB, and $\sigma_{\tt LI}^2=10$dB.
Figure~\ref{fig:8} illustrates the cumulative distributions of the
sum spectral efficiencies for the half-duplex, full-duplex, and
hybrid modes. The ZF processing outperforms the MRC/MRT processing
in this example, and the sum spectral efficiency of MRC/MRT
processing is more concentrated around its mean compared to the ZF
processing. Furthermore, we can see that, for MRC/MRT, the
full-duplex mode is always better than the half-duplex mode, while
for ZF, depending on the large-scale fading, full-duplex can be
better than half-duplex relaying and vice versa. In this example,
it is also shown that relaying using the hybrid mode provides a
large gain for the ZF processing case.

\begin{figure}[t]
    \centerline{\includegraphics[width=0.48\textwidth]{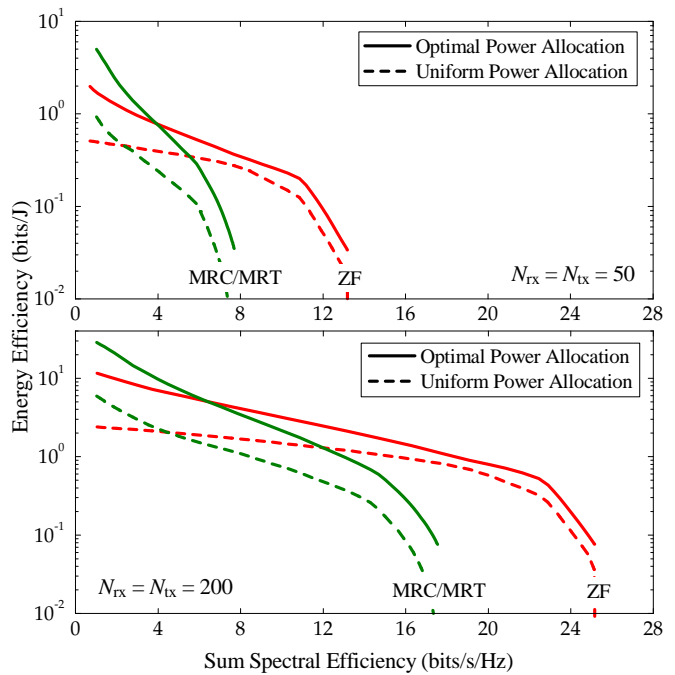}}
    \caption{Energy efficiency versus sum spectral efficiency for ZF and MRC/MRT ($K=10$, $\tau=2K$, $\Pp=10$dB, and $\sigma_{\tt LI}^2=10$dB).}
    \label{fig:9}
\end{figure}

\subsection{Power Allocation}
In the following, we will examine the energy efficiency versus the
sum spectral efficiency under the optimal power allocation, as
outlined in Section~\ref{sec: PA}. In this example, we choose
$\Pp=10$dB and $\sigma_{\tt LI}^2 = 10$dB. Furthermore, the
large-scale fading matrices are chosen as follows:
\begin{align*}
    \B{D}_{\tt SR}
        &=
        \mathrm{diag}\left[0.749  ~  0.246 ~   0.125 ~   0.635 ~   4.468 \right. \\ &\hspace{3.5cm}\left. 0.031 ~   0.064 ~   0.257 ~   0.195 ~   0.315 \right],
     \\
    \B{D}_{\tt RD}
        &=
        \mathrm{diag}\left[0.070 ~  0.121  ~  0.134 ~   0.209 ~   0.198  \right. \\ &\hspace{3.5cm}\left. 0.184 ~   0.065  ~  0.051 ~   0.236   ~ 1.641  \right].
\end{align*}
Note that, the above large-scale coefficients are obtained by
taking one snapshot of the practical setup for Fig.~\ref{fig:8}.

Figure~\ref{fig:9} shows the energy efficiency versus sum the
spectral efficiency under uniform and optimal power allocation.
The ``uniform power allocation" curves correspond to the case
where all sources and the relay station use their maximum powers,
i.e., $p_{{\tt S},k}=p_0$, $\forall k =1,..., K$, and $\PR=p_1$.
The ``optimal power allocation'' curves are obtained by using the
optimal power allocation scheme via Algorithm~\ref{sec: algo}. The
initial values of Algorithm~\ref{sec: algo} are chosen as follows:
$\epsilon=0.01$, $L=5$, $\alpha=1.1$, and $\gamma_{k,1}=\min
\left\{\frac{a_k p_0}{p_0\sum_{j=1}^K b_j  + c_k p_1 +1 },
\frac{d_k p_1}{e_k p_1 + 1}
        \right\}$ which correspond to the uniform power
        allocation case. We can see that with optimal power allocation, the system performance
improves significantly, especially at low spectral efficiencies.
For example, with $\Nr= \Nt= 200$, to achieve the same sum
spectral efficiency of $10$bits/s/Hz, optimal power allocation can
improve the energy efficiency by factors of $2$ and $3$ for ZF and
MRC/MRT processing, respectively, compared to the case of no power
allocation. This manifests that MRC/MRT processing benefits more
from power allocation. Furthermore, at low spectral efficiencies,
MRC/MRT performs better than ZF and vice versa at high spectral
efficiencies. The results also demonstrate the significant benefit
of using large antenna arrays at the relay station. With ZF
processing, by increasing the number of antennas from $50$ to
$200$, the energy efficiency can be increased by $14$ times, when
each pair has a throughput of about one bit per channel use.

\section{Conclusion} \label{Sec:Conclusion}
In this paper, we introduced and analyzed a multipair full-duplex
relaying system, where the relay station is equipped with massive
arrays, while each source and destination have a single antenna.
We assume that the relay station employs ZF and MRC/MRT to process
the signals. The analysis takes the energy and bandwidth costs of
channel estimation into account. We show that, by using massive
arrays at the relay station, loop interference can be canceled
out. Furthermore, the interpair interference and noise disappear.
As a result, massive MIMO can increase the sum spectral efficiency
by $2K$ times compared to the conventional orthogonal half-duplex
relaying, and simultaneously reduce the transmit power
significantly. We derived closed-form expressions for the
achievable rates and compared the performance of the full-duplex
and half-duplex modes. In addition, we proposed a power allocation
scheme which  chooses optimally the transmit powers of the $K$
sources and of the relay station  to maximize the energy
efficiency, subject to a given sum spectral efficiency and peak
power constraints. With the optimal power allocation, the energy
efficiency can be significantly improved.

\appendix

\subsection{Proof of Proposition~\ref{PropLarge1}} \label{Appendix Large1}

\begin{enumerate}

\item For ZF processing:

Here, we first provide the proof for ZF processing. From \eqref{eq
MMSECE 2} and \eqref{eq ZFReceiver 1}, we have
\begin{align}\label{eq large 1}
    \sqrt{\Ps}
    \B{W}^T
    \B{G}_{\tt SR}
    \B{x}\left[i\right]
    &\!=\!
    \sqrt{\Ps}
    \B{W}_{\tt ZF}^T
    \left(
        \hat{\B{G}}_{\tt SR}
        +
        \pmb{\mathcal{E}}_{\tt SR}
    \right)
    \B{x}\left[i\right]\nonumber\\
    &\!=\!
    \sqrt{\Ps}\B{x}\left[i\right]
    \!+\!
    \sqrt{\Ps}\B{W}_{\tt ZF}^T\pmb{\mathcal{E}}_{\tt
    SR}\B{x}\left[i\right].
\end{align}
By using the law of large numbers, we obtain\footnote{
    The law of large numbers: Let $\B{p}$
and $\B{q}$ be mutually independent $n \times 1$ vectors. Suppose
that the elements of $\B{p}$ are i.i.d. zero-mean random variables
with variance $\sigma_{p}^2$, and that the elements of $\B{q}$ are
i.i.d. zero-mean random variables  with variance $\sigma_{q}^2$.
Then, we have\begin{align*}
    \frac{1}{n} \B{p}^H \B{p} \mathop  \to \limits^{a.s.} \sigma_p^2, ~
    \text{and} ~ ~
    \frac{1}{n} \B{p}^H \B{q} \mathop  \to \limits^{a.s.} 0, ~ \text{as} ~
    n \rightarrow \infty.
\end{align*}
}
\begin{align}\label{eq large 2}
    \sqrt{\Ps}
    \B{W}_{\tt ZF}^T\pmb{\mathcal{E}}_{\tt SR}\B{x}\left[i\right]
    &=
    \sqrt{\Ps}
    \left(\frac{\hat{\B{G}}_{\tt SR}^H \hat{\B{G}}_{\tt SR}}{\Nr}\right)^{-1}
    \frac{\hat{\B{G}}_{\tt SR}^H
    \pmb{\mathcal{E}}_{\tt
    SR}}{\Nr}\B{x}\left[i\right]\nonumber\\&
    \mathop  \to \limits^{a.s.}
    0, ~ \text{as} ~ \Nr\to\infty.
\end{align}
Therefore, as $\Nr\to\infty$, we have
\begin{align}\label{eq large 3}
    \sqrt{\Ps}
    \B{W}^T
    \B{G}_{\tt SR}
    \B{x}\left[i\right]
    \mathop  \to \limits^{a.s.}
    \sqrt{\Ps}\B{x}\left[i\right].
\end{align}
From \eqref{eq large 3}, we can see that, when $\Nr$ goes to
infinity, the desired signal converges to a deterministic value,
while the multi-pair interference is cancelled out. More
precisely, as $\Nr\to\infty$,
\begin{align}\label{eq large 4}
    \sqrt{\Ps}
    \B{w}_k^T
    \B{g}_{{\tt SR},k}
    {x}_k\left[i\right]
    &\mathop  \to \limits^{a.s.}
    \sqrt{\Ps}{x}_k\left[i\right],
    \\
    \sqrt{\Ps}
    \B{w}_k^T
    \B{g}_{{\tt SR},j}
    {x}_j\left[i\right]
    &\mathop  \to \limits^{a.s.}
    0, ~ \forall j \neq k. \label{eq large 4b}
\end{align}

Next, we consider the loop interference. With ZF processing, we
have
\begin{align} \label{eq large 5}
    \sqrt{\PR}
    \B{W}^T
    \B{G}_{\tt RR}
    \B{s}\left[i\right]
    &=
    \alpha_{\tt ZF}
    \sqrt{\PR}
    \left(\frac{\hat{\B{G}}_{\tt SR}^H \hat{\B{G}}_{\tt
    SR}}{\Nr}\right)^{-1}\nonumber\\ &\!\hspace{-1cm}\times\!
    \frac{\hat{\B{G}}_{\tt SR}^H
    \B{G}_{\tt RR} \hat{\B{G}}_{\tt RD}^{\ast}}{\Nr\Nt}
    \left(\!
    \frac{
    \hat{\B{G}}_{\tt RD}^T
    \hat{\B{G}}_{\tt RD}^{\ast}}{\Nt}
    \!\right)^{-1}\!\!\!
    \B{x}\left[i\!-\!d\right].
\end{align}
If $\Nt$ is fixed, then it is obvious that $    \sqrt{\PR}
    \B{W}^T
    \B{G}_{\tt RR}
    \B{s}\left[i\right] \to 0$, as $\Nr\to\infty$. We now consider
the case where $\Nt$ and $\Nr$ tend to infinity with a fixed
ratio. The $(m,n)$th element of the $K\times K$ matrix $
\alpha_{\tt ZF}\frac{\hat{\B{G}}_{\tt SR}^H
    \B{G}_{\tt RR} \hat{\B{G}}_{\tt RD}^{\ast}}{\Nr\Nt}$ can be
    written as
\begin{align} \label{eq large 6}
    \alpha_{\tt ZF}\!\frac{\hat{\B{g}}_{{\tt SR},m}^H\!
    \B{G}_{\tt RR} \hat{\B{g}}_{{\tt RD},n}^{\ast}}{\Nr\Nt}
    \!\!=\!\!
    \sqrt{
        \frac{
            \Nt-K
            }{
            \Nt\!\!\sum\limits_{k=1}^K\!\!\sigma_{{\tt RD},k}^{-2}
            }
    }
    \frac{1}{\Nr}\!
    \hat{\B{g}}_{{\tt SR},m}^H\!
    \frac{\B{G}_{\tt RR} \hat{\B{g}}_{{\tt
    RD},n}^{\ast}}{\sqrt{\Nt}}.
\end{align}
We can see that the vector $\frac{\B{G}_{\tt RR} \hat{\B{g}}_{{\tt
RD},n}^{\ast}}{\sqrt{\Nt}}$ includes i.i.d. zero-mean random
variables with variance $\sigma_{{\tt RD},n}^{2}\sigma_{\tt
LI}^2$. This vector is independent of the vector
$\hat{\B{g}}_{{\tt SR},m}$. Thus, by using the law of large
numbers, we can obtain
\begin{align} \label{eq large 6}
    \alpha_{\tt ZF}\!\frac{\hat{\B{g}}_{{\tt SR},m}^H
    \!\B{G}_{\tt RR} \hat{\B{g}}_{{\tt RD},n}^{\ast}}{\Nr\Nt}
    \mathop  \to \limits^{a.s.} 0, ~ \text{as} ~ \Nr\!\to\!\infty, \Nr/\Nt ~ \text{is fixed}.
\end{align}
Therefore, the loop interference converges to $0$ when $\Nr$ grows
without bound. Similarly, we can show that
\begin{align} \label{eq large 7}
    \B{W}^T
    \B{n}_{\tt R}\left[i\right]
    \mathop  \to \limits^{a.s.}
    0.
\end{align}
Substituting \eqref{eq large 4}, \eqref{eq large 4b}, \eqref{eq
large 6}, and \eqref{eq large 7} into \eqref{eq ZFR 1b}, we arrive
at \eqref{eq PropLarge 1}.

\item For MRC/MRT processing:

We next provide the proof for MRC/MRT processing.  From \eqref{eq
MMSECE 2} and \eqref{eq MRCReceiver 1}, and by using the law of
large numbers, as $\Nr\to\infty$, we have that
\begin{align}\label{eq large1 4}
    \frac{1}{\Nr}\sqrt{\Ps}
    \B{w}_k^T
    \B{g}_{{\tt SR},k}
    {x}_k\left[i\right]
    &=
    \frac{1}{\Nr}\sqrt{\Ps}
    \hat{\B{g}}_{{\tt SR},k}^H
    \B{g}_{{\tt SR},k}
    {x}_k\left[i\right]\nonumber\\
    &\mathop  \to \limits^{a.s.}
    \sqrt{\Ps}\sigma_{{\tt SR},k}^2{x}_k\left[i\right],
    \\
    \frac{1}{\Nr}\sqrt{\Ps}
    \B{w}_K^T
    \B{g}_{{\tt SR},j}
    {x}_k\left[j\right]
    &=
    \frac{1}{\Nr}\sqrt{\Ps}
    \hat{\B{g}}_{{\tt SR},k}^H
    \B{g}_{{\tt SR},j}
    {x}_k\left[j\right]\nonumber\\
    &
    \mathop  \to \limits^{a.s.}
    0, ~ \forall j \neq k. \label{eq large1 4b}
\end{align}
We next consider the loop interference. For any
finite $\Nt$, or any $\Nt$ where $\Nr/\Nt$ is fixed, as
$\Nr\to\infty$, we have
\begin{align} \label{eq large1 5}
    \frac{1}{\Nr}\sqrt{\PR}
    \B{W}^T
    \B{G}_{\tt RR}
    \B{s}\left[i\right]
    &=
    \alpha_{\tt MRT}
    \sqrt{\PR}
    \frac{\hat{\B{G}}_{\tt SR}^H
    \B{G}_{\tt RR} \hat{\B{G}}_{\tt RD}^{\ast}}{\Nr}
    \B{x}\left[i-d\right]\nonumber\\
    &
    \mathop  \to \limits^{a.s.} 0,
\end{align}
where the convergence follows  a
similar argument as in the proof for ZF processing. Similarly, we
can show that
\begin{align} \label{eq large1 7}
    \frac{1}{\Nr}\B{w}_k^T
    \B{n}_{\tt R}\left[i\right]
    \mathop  \to \limits^{a.s.} 0.
\end{align}
Substituting \eqref{eq large1 4}, \eqref{eq large1 4b}, \eqref{eq
large1 5}, and \eqref{eq large1 7} into \eqref{eq ZFR 1b}, we
obtain  \eqref{eq PropLarge 1b}.

\end{enumerate}

%

\subsection{Proof of Theorem~\ref{Theorem 1}} \label{sec app thrm1}
\subsubsection{Derive $R_{{\tt SR}_k}$} From \eqref{eq Rate1
1}, we need to compute $\E\left\{\B{w}_k^T \B{g}_{{\tt SR},k}
\right\}$, $\var\left(\B{w}_k^T \B{g}_{{\tt SR},k}\right)$, ${\tt
MP}_k$, ${\tt LI}_k$, and ${\tt AN}_k$.
\begin{itemize}
\item Compute $\E\left\{\B{w}_k^T \B{g}_{{\tt SR},k} \right\}$:

Since, $\B{W}^T = \left(\hat{\B{G}}_{\tt SR}^H \hat{\B{G}}_{\tt
SR}\right)^{-1}\hat{\B{G}}_{\tt SR}^H$, from \eqref{eq MMSECE 2},
we have
\begin{align} \label{eq Proof1 20}
    \B{W}^T{\B{G}}_{\tt SR}
    =
    \B{W}^T\left(\hat{\B{G}}_{\tt SR} + \pmb{\mathcal{E}}_{\tt SR} \right)
    =
    \B{I}_{\Nr} + \B{W}^T\pmb{\mathcal{E}}_{\tt SR}.
\end{align}
Therefore,
\begin{align} \label{eq Proof1 2}
    \B{w}_k^T{\B{g}}_{{\tt SR},k}
    =
    1 + \B{w}_k^T\pmb{\varepsilon}_{{\tt SR},k}
\end{align}
where $\pmb{\varepsilon}_{{\tt SR},k}$ is the $k$th column of
$\pmb{\mathcal{E}}_{\tt SR}$. Since $\pmb{\varepsilon}_{{\tt
SR},k}$ and $\B{w}_k$ are uncorrelated, and
$\pmb{\varepsilon}_{{\tt SR},k}$ is a zero-mean random variable,
$\E\left\{
        \B{w}_k^T \pmb{\varepsilon}_{{\tt SR},k}
    \right\} = 0$. Thus,
\begin{align} \label{eq Proof1 2b}
    \E\left\{\B{w}_k^T \B{g}_{{\tt SR},k} \right\}
    =
    1.
\end{align}

\item Compute $\var\left(\B{w}_k^T \B{g}_{{\tt SR},k}\right)$:

From \eqref{eq Proof1 2} and \eqref{eq Proof1 2b}, the variance of
$\B{w}_k^T \B{g}_{{\tt SR},k}$ is given by
\begin{align} \label{eq Proof1 3}
    \var\left(\B{w}_k^T \B{g}_{{\tt SR},k}\right)
    &=
    \E\left\{
        \left|\B{w}_k^T \pmb{\varepsilon}_{{\tt SR},k}\right|^2
    \right\}
     \nonumber
     \\
     &=
    \left(\beta_{{\tt SR},k} - \sigma_{{\tt SR},k}^2\right)\E\left\{
        \left\|\B{w}_k\right\|^2
    \right\}
    \nonumber
    \\
    &=
    \left(\beta_{{\tt SR},k} \!- \!\sigma_{{\tt SR},k}^2\right)\!
    \E\left\{\!
        \left[\left(\hat{\B{G}}_{\tt SR}^H \hat{\B{G}}_{\tt
        SR}\right)^{-1}\right]_{kk}
     \!\right\}
     \nonumber
     \\
     &=
     \frac{\beta_{{\tt SR},k} - \sigma_{{\tt SR},k}^2}{\sigma_{{\tt SR},k}^2 K}
     \E\left\{\tr
        \left(\B{X}^{-1}\right)
     \right\}
    \nonumber
    \\
    &=
    \frac{\beta_{{\tt SR},k} - \sigma_{{\tt SR},k}^2}{\sigma_{{\tt SR},k}^2 }
     \frac{1}{\Nr-K}, ~ \text{for} ~ \Nr > K,
\end{align}
where $\B{X}$ is a $K\times K$ central Wishart matrix with $\Nr$
degrees of freedom and covariance matrix $\B{I}_K$, and the last
equality is obtained by using \cite[Lemma~2.10]{TV:04:FTCIT}.

\item Compute ${\tt MP}_k$:

From \eqref{eq Proof1 20}, we have that $ \B{w}_k^T{\B{g}}_{{\tt
SR},j}
    =
    \B{w}_k^T\pmb{\varepsilon}_{{\tt SR},j}$, for $j \neq k$. Since
    $\B{w}_k$ and $\pmb{\varepsilon}_{{\tt SR},j}$ are uncorrelated,
    we obtain
\begin{align} \label{eq Proof1 4}
    \E\left\{
        \left|\B{w}_k^T \pmb{\varepsilon}_{{\tt SR},j}\right|^2
    \right\}
    &=
    \left(\beta_{{\tt SR},j} - \sigma_{{\tt SR},j}^2\right)
    \E\left\{
        \left\|\B{w}_k\right\|^2
    \right\}
        \nonumber
    \\
    &=
    \frac{\beta_{{\tt SR},j} - \sigma_{{\tt SR},j}^2}{\sigma_{{\tt SR},k}^2 }
     \frac{1}{\Nr-K}.
\end{align}
Therefore,
\begin{align} \label{eq Proof1 4b}
    {\tt MP}_k
    =
    \Ps
    \sum_{j \neq K}^K
    \frac{\beta_{{\tt SR},j} - \sigma_{{\tt SR},j}^2}{\sigma_{{\tt SR},k}^2 }
     \frac{1}{\Nr-K}.
\end{align}

\item Compute ${\tt LI}_k$:

From \eqref{eq Rate1 2b}, with ZF, the LI can be rewritten as
\begin{align}\label{eq Rate1 3}
    {\tt LI}_k
    &=
    \PR
    \E
    \left\{
        \B{w}_k^T
        \B{G}_{\tt RR}
        \B{A}_{\tt ZF}\B{A}_{\tt ZF}^H \B{G}_{\tt RR}^H \B{w}_k^\ast
    \right\}.
\end{align}
From \eqref{eq ZFPrecoder 1}, we have
\begin{align}\label{eq Rate1 4}
    \B{A}_{\tt ZF}\B{A}_{\tt ZF}^H
    =
    \alpha_{\tt ZF}^2
    \hat{\B{G}}_{\tt RD}^{\ast}\left(\hat{\B{G}}_{\tt RD}^T \hat{\B{G}}_{\tt RD}^{\ast}\right)^{-1}
    \left(\hat{\B{G}}_{\tt RD}^T \hat{\B{G}}_{\tt RD}^{\ast}\right)^{-1}\hat{\B{G}}_{\tt
    RD}^T.
\end{align}
When $\Nt \gg K$, we can use the law of large numbers to obtain
the following approximation:
\begin{align}\label{eq Rate1 5}
    \hat{\B{G}}_{\tt RD}^T \hat{\B{G}}_{\tt RD}^{\ast}
    \approx
    \Nt \hat{\B{D}}_{\tt RD},
\end{align}
where $\hat{\B{D}}_{\tt RD}$ is a $K\times K$ diagonal matrix
whose $(k, k)$th element is $\left[\hat{\B{D}}_{\tt
RD}\right]_{kk}=\sigma_{{\tt RD},k}^2$. Therefore,
\begin{align}\label{eq Rate1 6}
    \B{A}_{\tt ZF}\B{A}_{\tt ZF}^H
    \approx
    \frac{\alpha_{\tt ZF}^2}{\Nt^2 }
    \hat{\B{G}}_{\tt RD}^{\ast}  \hat{\B{D}}_{\tt RD}^{-2} \hat{\B{G}}_{\tt
    RD}^T.
\end{align}
Substituting \eqref{eq Rate1 6} into \eqref{eq Rate1 3} we obtain

\begin{align}\label{eq Proof1 5}
    {\tt LI}_k
    &\approx
    \PR
    \frac{\alpha_{\tt ZF}^2}{\Nt^2}
    \E
    \left\{
        \B{w}_k^T
        \B{G}_{\tt RR}
        \hat{\B{G}}_{\tt RD}^{\ast}  \hat{\B{D}}_{\tt RD}^{-2} \hat{\B{G}}_{\tt RD}^T
        \B{G}_{\tt RR}^H \B{w}_k^\ast
    \right\}
    \nonumber
    \\
    &=
    \PR
    \frac{\alpha_{\tt ZF}^2}{\Nt^2 }
    \left(\sum_{j=1}^K \frac{1}{\sigma_{{\tt RD},j}^2}\right)
    \E
    \left\{
        \B{w}_k^T
        \B{G}_{\tt RR}
        \B{G}_{\tt RR}^H \B{w}_k^\ast
    \right\}
    \nonumber
    \\
    &=
    \PR
    \frac{\alpha_{\tt ZF}^2\sigma_{\tt LI}^2}{\Nt }
    \left(\sum_{j=1}^K \frac{1}{\sigma_{{\tt RD},j}^2}\right)
    \E
    \left\{
        \left\|
        \B{w}_k
        \right\|^2
    \right\}
        \nonumber
    \\
    &=
    \frac{\sigma_{\tt LI}^2\PR\left(\Nt-K\right)}{\sigma_{{\tt SR},k}^2 \Nt \left(\Nr-K
    \right)}.
\end{align}
\item Compute ${\tt AN}_k$:

Similarly, we obtain
\begin{align}\label{eq Proof1 6}
    {\tt AN}_k
    =
    \frac{1}{\sigma_{{\tt SR},k}^2} \frac{1}{\Nr-K}.
\end{align}

\end{itemize}

Substituting \eqref{eq Proof1 2b}, \eqref{eq Proof1 3}, \eqref{eq
Proof1 4b}, \eqref{eq Proof1 5}, and \eqref{eq Proof1 6} into
\eqref{eq Rate1 1}, we obtain
\begin{align} \label{eq Proof1 7}
    R_{{\tt SR}, k}
    \!\approx\!
    \log_2\!
    \left(\!
        1
        \!+\!
        \frac{
            \Ps
            \left(\Nr-K \right) \sigma_{{\tt SR},k}^2
            }{
            \Ps \sum\limits_{j=1}^K \!\!\left(\beta_{{\tt SR},j}\!-\! \sigma_{{\tt SR},j}^2 \!\right)
            \!+\!
            \PR\!\sigma_{\tt LI}^2\left(\!1\!-\! \frac{K}{\Nt}\! \right)
            \!+\!
            1
            }
    \right).
\end{align}

\subsubsection{Derive $R_{{\tt RD}, k}$}
From \eqref{eq LPRate1 2}, to derive $R_{{\tt RD}, k}$, we need to
compute $\E\left\{\B{g}_{{\tt RD},k}^T\B{a}_k \right\}$,
$\var\left(\B{g}_{{\tt RD},k}^T\B{a}_k\right)$, and
$\E\left\{\left|
                \B{g}_{{\tt RD},k}^T\B{a}_j
            \right|^2\right\}$. Following the same methodology as the one used to
            compute $\E\left\{\B{w}_k^T \B{g}_{{\tt SR},k}
\right\}$, $\var\left(\B{w}_k^T \B{g}_{{\tt SR},k}\right)$, and
${\tt MP}_k$, we obtain
\begin{align} \E\left\{\B{g}_{{\tt
RD},k}^T\B{a}_k \right\} &= \alpha_{\tt
ZF},\label{eq Proof1rd 1} \\
\var\left(\B{g}_{{\tt RD},k}^T\B{a}_k\right) &=
\frac{\left(\beta_{{\tt RD},k}-\sigma_{{\tt
RD},k}^2\right)\alpha_{\tt ZF}^2}{\sigma_{{\tt
RD},k}^2\left(\Nt-K\right)}, \label{eq Proof1rd 2} \\
\E\left\{\left|
                \B{g}_{{\tt RD},k}^T\B{a}_j
            \right|^2\right\} &=
\frac{\left(\beta_{{\tt RD},k}-\sigma_{{\tt
RD},k}^2\right)\alpha_{\tt ZF}^2}{\sigma_{{\tt
RD},j}^2\left(\Nt-K\right)}, ~ \text{for} ~ j\neq k.\label{eq
Proof1rd 3}
\end{align}
Substituting \eqref{eq Proof1rd 1}--\eqref{eq Proof1rd 3} into
\eqref{eq LPRate1 2}, we obtain a closed-form expression for
$R_{{\tt RD},k}$:
\begin{align} \label{eq RateRD 1}
    R_{{\tt RD},k}
    =
    \log_2
    \left(
        1
        +
        \frac{
            \Nt - K
            }{
            \sum_{j=1}^K \sigma_{{\tt RD}, j}^{-2}
            }
        \frac{
            \PR
            }{
            \PR\left(\beta_{{\tt RD},k}- \sigma_{{\tt RD},k}^2 \right) + 1
            }
    \right).
\end{align}
Then, using \eqref{eq RateAnalysis
1}, \eqref{eq Proof1 7}, and \eqref{eq RateRD 1}, we arrive at
\eqref{eq Rate2 re}.

\subsection{Proof of Theorem~\ref{Theorem 2}} \label{sec app thrm2}
With MRC/MRT processing,  $\B{W}^T = \hat{\B{G}}_{\tt SR}^H$ and
$\B{A} =\alpha_{\tt MRT} \hat{\B{G}}_{\tt RD}^{\ast}$.

\begin{enumerate}
\item Compute $\E\left\{\B{w}_k^T \B{g}_{{\tt SR},k} \right\}$:

We have
\begin{align} \label{eq ProofMRC 1}
    \B{w}_k^T {\B{g}}_{{\tt SR},k}
    =
    \hat{\B{g}}_{{\tt SR},k}^H \B{g}_{{\tt SR},k}
    =
    \left\|\hat{\B{g}}_{{\tt SR},k}\right\|^2 + \hat{\B{g}}_{{\tt SR},k}^H\pmb{\varepsilon}_{{\tt SR},k}.
\end{align}
Therefore,
\begin{align} \label{eq ProofMRC 2b}
    \E\left\{\B{w}_k^T \B{g}_{{\tt SR},k} \right\}
    =
    \E\left\{\left\|\hat{\B{g}}_{{\tt SR},k}\right\|^2 \right\}
    =\sigma_{{\tt SR}, k}^2\Nr.
\end{align}

\item Compute $\var\left(\B{w}_k^T \B{g}_{{\tt SR},k}\right)$:

From \eqref{eq ProofMRC 1} and \eqref{eq ProofMRC 2b}, the
variance of $\B{w}_k^T \B{g}_{{\tt SR},k}$ is given by
\begin{align} \label{eq ProofMRC 3}
    &\var\left(\B{w}_k^T \B{g}_{{\tt SR},k}\right)
    =
    \E\left\{
        \left|\B{w}_k^T \B{g}_{{\tt SR},k}\right|^2
    \right\}
    -
    \sigma_{{\tt SR}, k}^4\Nr^2
    \nonumber
    \\
    &\hspace{-0cm}=
    \E\left\{
        \left| \left\|\hat{\B{g}}_{{\tt SR},k}\right\|^2 + \hat{\B{g}}_{{\tt SR},k}^H\pmb{\varepsilon}_{{\tt SR},k}\right|^2
    \right\}
    -
    \sigma_{{\tt SR}, k}^4\Nr^2
    \nonumber
    \\
    &\hspace{-0cm}=
    \E\left\{
        \left\|\hat{\B{g}}_{{\tt SR},k}\right\|^4
    \right\}
    +
    \E\left\{
        \left|\hat{\B{g}}_{{\tt SR},k}^H\pmb{\varepsilon}_{{\tt SR},k}\right|^2
    \right\}
    -
    \sigma_{{\tt SR}, k}^4\Nr^2.
\end{align}
By using \cite[Lemma~2.9]{TV:04:FTCIT}, we obtain
\begin{align} \label{eq ProofMRC 3b}
    \var\left(\B{w}_k^T \B{g}_{{\tt SR},k}\right)
    &=
    \sigma_{{\tt SR}, k}^4 \Nr\left(\Nr+1 \right)
        \nonumber
    \\
    &\hspace{-0cm}+
    \sigma_{{\tt SR}, k}^2\left(\beta_{{\tt SR}, k} - \sigma_{{\tt SR}, k}^2 \right) \Nr
    -
    \sigma_{{\tt SR}, k}^4\Nr^2
    \nonumber
    \\
    &=\sigma_{{\tt SR}, k}^2 \beta_{{\tt SR}, k} \Nr.
\end{align}

\item Compute ${\tt MP}_k$:

For $j\neq k$, we have
\begin{align} \label{eq ProofMRC 4}
    \E\left\{
        \left|\B{w}_k^T \B{g}_{{\tt SR},j}\right|^2
    \right\}
    =
    \E\left\{
        \left|\hat{\B{g}}_{{\tt SR},k}^H \B{g}_{{\tt SR},j}\right|^2
    \right\}
    =
    \sigma_{{\tt SR}, k}^2 \beta_{{\tt SR}, j} \Nr.
\end{align}
Therefore,
\begin{align} \label{eq ProofMRC 4b}
    {\tt MP}_k
    =
    \Ps
    \sigma_{{\tt SR}, k}^2  \Nr \sum_{j \neq k}^K \beta_{{\tt SR}, j}.
\end{align}

\item Compute ${\tt LI}_k$:

Since $\hat{\B{g}}_{{\tt SR},k}$, $\B{G}_{\tt RR}$, and
$\hat{\B{G}}_{\tt RD}$ are independent, we obtain
\begin{align}\label{eq ProofMRC 5}
    {\tt LI}_k
    &=
    \alpha_{\tt MRT}^2\PR
    \E
    \left\{
        \hat{\B{g}}_{{\tt SR},k}^H
        \B{G}_{\tt RR}
        \hat{\B{G}}_{\tt RD}^\ast\hat{\B{G}}_{\tt RD}^T
        \B{G}_{\tt RR}^H
        \hat{\B{g}}_{{\tt SR},k}^\ast
    \right\}
    \nonumber
    \\
    &=
    \alpha_{\tt MRT}^2\PR
    \left(\sum_{j=1}^K \sigma_{{\tt RD}, j}^2\right)
    \E
    \Big\{
        \hat{\B{g}}_{{\tt SR},k}^H
        \B{G}_{\tt RR}
        \B{G}_{\tt RR}^H
        \hat{\B{g}}_{{\tt SR},k}^\ast
    \Big\}
    \nonumber
    \\
    &=
    \alpha_{\tt MRT}^2\PR
    \left(\sum_{j=1}^K \sigma_{{\tt RD}, j}^2\right)
    \sigma_{\tt LI}^2
    \Nt
    \E
    \Big\{
        \hat{\B{g}}_{{\tt SR},k}^H
        \hat{\B{g}}_{{\tt SR},k}^\ast
    \Big\}
        \nonumber
    \\
    &\hspace{-0cm}=
   \PR \sigma_{\tt LI}^2 \sigma_{{\tt SR}, k}^2 \Nr.
\end{align}
\item Compute ${\tt AN}_k$:

Similarly, we obtain
\begin{align}\label{eq ProofMRC 6}
    {\tt AN}_k
    =
    \sigma_{{\tt SR}, k}^2 \Nr.
\end{align}

\end{enumerate}

Substituting \eqref{eq ProofMRC 2b}, \eqref{eq ProofMRC 3b},
\eqref{eq ProofMRC 4b}, \eqref{eq ProofMRC 5}, and \eqref{eq
ProofMRC 6} into \eqref{eq Rate1 1}, we obtain
\begin{align} \label{eq ProofMRC 7}
    R_{{\tt SR}, k}
    =
    \log_2
    \left(
        1
        +
        \frac{
            \Ps
            \Nr \sigma_{{\tt SR},k}^2
            }{
            \Ps \sum_{j=1}^K \beta_{{\tt SR},j}
            +
            \PR\sigma_{\tt LI}^2
            +
            1
            }
    \right).
\end{align}
Similarly, we obtain a closed-form expression for $R_{{\tt RD},
k}$, and then we arrive at \eqref{eq RateMRC 1}.



\end{document}